\documentclass[fleqn,usenatbib]{mnras}

\usepackage[T1]{fontenc}
\usepackage{ae,aecompl}

\usepackage{graphicx, float}	
\usepackage{amsmath}	
\usepackage{amssymb}	
\usepackage[hyphenbreaks]{breakurl}

\title[Spectral and Temporal Properties of NGC~55~ULX1]{Spectral and Temporal Properties of Ultra-luminous X-ray Source NGC~55~ULX1}

\author[Jithesh]{
V. Jithesh$^{1,2}$\thanks{E-mail: jitheshthejus@gmail.com}\\
$^{1}$Inter-University Centre for Astronomy and Astrophysics (IUCAA), PB No.4, Ganeshkhind, Pune-411007, India\\
$^{2}$Department of Physics, University of Calicut, Malappuram, Kerala-673635, India\\ 
}

\date{Accepted 2021 November 11. Received 2021 November 11; in original form 2021 October 20}
\pubyear{2021}

\begin{document}
\label{firstpage}
\pagerange{\pageref{firstpage}--\pageref{lastpage}}
\maketitle

\begin{abstract}

We investigate the spectral and temporal properties of ultra-luminous X-ray source (ULX) NGC~55~ULX1 using {\it Swift}, {\it XMM-Newton} and {\it NuSTAR} observations conducted during 2013--2021. In these observations, the source flux varies by a factor of $\sim 5-6$, and we identify the source mainly in the $soft~ultraluminous$ (SUL) state of ULXs. We fit the X-ray spectra with a two thermal component model consisting of a blackbody (for the soft component) and a disc (for the hard component), and the soft component dominates in these observations. The soft component in the SUL state shows properties similar to that of ultraluminous supersoft sources, for example, an anti-correlation between the characteristic radius and temperature of the blackbody component. In addition, we observe a positive correlation between the blackbody and inner disc temperatures when the X-ray spectra are fitted with the two-thermal component model. The source exhibits marginal evidence of X-ray flux dips in the {\it Swift} and {\it XMM-Newton} observations at different intensity levels. We explain the observed spectral and temporal properties of the source by invoking the supercritical radiatively driven outflow mechanism.
\end{abstract}




\section{Introduction}
\label{sec:intro}
 
Ultra-luminous X-ray sources (ULXs) are bright, non-nuclear accreting X-ray sources with an isotropic X-ray luminosity above $10^{39}\rm erg~s^{-1}$ \citep[see][for a recent review]{Kaa17}. The inferred high luminosity is considered as strong evidence for the existence of intermediate-mass black hole (IMBH) in ULXs with a proposed mass range of $10^{2}-10^{5} M_{\odot}$ \citep{Col99, Mil04a}. The detailed studies in the last couple of decades with {\it XMM-Newton} observations do not identify any signatures of the sub-Eddington accretion flow onto the IMBH, instead, these studies favour supercritical accretion onto stellar-mass black hole \citep[e.g.,][]{Gla09, Mid12, Sut13, Mid15a, Gho21}. The radiatively driven optically-thick outflows expected in the super-Eddington accretion systems \citep{Pou07, Mid15a} play an important role in understanding the X-ray spectral and temporal properties of these sources. The recent detection of pulsations from several ULXs confirms the presence of neutron star (NS) compact object in the ULX population \citep{Bac14, 2017Sci...355..817I, 2017MNRAS.466L..48I, Car18, Bri18, Sat19, Rod20}. Pulsations are detected when the pulsed fraction and count rates are high. However, in the absence of pulsations, the broadband spectral analysis provided an alternate way to understand the compact object and identified several candidate pulsar ULXs \citep{Pin17, Wal18, Jit20, Jit21, Gur21}. The compact object of ULXs is still under debate, and recent studies suggest a heterogeneous population comprised of NS, stellar-mass black hole and IMBH in some extreme cases \citep[for example][]{Far09, Fen10, Ser11, Pas14}. 

Spectral analysis of high-quality {\it XMM-Newton} spectra suggested two components for ULXs: the soft excess and the high energy curvature \citep[e.g.,][]{Sto06, Gla09, Sut13}. The hard emission could arise either from the innermost regions of the accretion flow or from inverse-Compton scattering of seed photons from the inner disc in a hot corona \citep{Rob07, Mid15a, Muk15, Jit17}, while the soft component can be interpreted either as the thermal emission associated with an outflowing wind \citep{Pou07, Mid15a} or as the emission from the disc \citep{Mil13}. Using the two-component modelling, \citet{Sut13} empirically classified ULXs into three spectral regimes, $broadened~disc$, $hard~ultraluminous$ (HUL) and $soft~ultraluminous$ (SUL) based on the spectral morphologies observed in the X-ray spectrum. They argued that the spectral and temporal properties of ULXs are consistent with the super-Eddington accretion model, and the observed properties can be explained by a massive radiatively driven wind, which forms a funnel-like geometry around the accretion flow. In addition, the key factors that determine the appearance of spectral characteristics are inclination and mass accretion rate \citep{Mid11a,  Sut13, Mid14}. Many soft ULXs showed X-ray residuals in the soft energy band ($< 2$ keV) and the flux drop around 1 keV in their X-ray CCD spectra. In addition, these ULXs exhibited anti-correlation between the temperature and radius. These features are believed to be the signatures of outflowing wind from the system \citep{Sut15, Fen16, Sor16, Urq16, Pin20}. Recently, the direct detection of outflowing wind has been reported in a handful of ULXs \citep{Pin16, Pin17, Pin21}, which confirm the existence of the outflow from super-Eddington accretion systems. 

Ultraluminous supersoft sources (ULSs) are the special subclass of ULXs, which have a dominant thermal X-ray emission below 1 keV with little or no emission at higher energies \citep{Urq16, Sor16}. The strong thermal component of ULS spectra is usually modelled with a blackbody model. The blackbody temperature $\rm kT_{bb} \sim$ 50--150 eV, the characteristic blackbody radius $\rm R_{bb} \sim 10^{4}-10^{5}$ km and the bolometric luminosity reaches $\approx$ a few $10^{39}\rm~erg~s^{-1}$ in the ULS state. Three different interpretations have been suggested for the soft component in ULSs: (1) nuclear burning on the surface of white dwarfs; (2) accretion disc emission from IMBH in the high/soft state; (3) a massive radiatively driven outflow by super-Eddington accreting stellar-mass black holes or NSs \citep[][and references therein]{Urq16}. The characteristic blackbody radius, the X-ray luminosity and the Eddington limit inferred from the ULSs spectra are not self-consistent with the accreting white dwarf scenario and IMBH interpretation. However, we cannot completely rule out these interpretations for ULSs. Another interesting feature detected in ULXs is the X-ray flux dipping \citep{Sto04, Liu05, Lin13, Pas13, Fen16, Urq16, Ai21, Als21, Hu21}. X-ray dips are not generally periodic and lasted for a few hundreds of seconds to several ks. In addition, the dips are often associated with the change in spectral hardness \citep{Lin13, Fen16, Ai21} however, such a change is not seen in some ULXs \citep{Pas13, Hu21}. These dips can be explained by different mechanisms such as the absorption effect by the structures (accretion stream-disc interaction region or the companion star's atmosphere) of the binary system, the geometrical occultation of the emitting regions, propeller effect and outflows driven by super-critical accretion \citep{Sto04, Fen16, Ai21, Als21, Hu21}. However, the exact mechanism responsible for X-ray dips in ULXs is still under debate.     

NGC~55~ULX1 is a bright, non-nuclear X-ray source in the Magellanic-type barred spiral galaxy NGC 55, located at a distance of 1.78 Mpc \citep{Kar03}, with a peak X-ray luminosity of $\sim 2 \times 10^{39}\rm erg~s^{-1}$. The source showed energy-dependent X-ray dips in the {\it XMM-Newton} observations conducted in 2001, and these dips have lasted for a few hundred seconds in these observations \citep{Sto04}. In the {\it Chandra} and {\it Swift} observations conducted during the period 2001--2013, the source exhibited marginal evidence of dip episodes, although the counting statistics of these data sets were poor \citep{Pin15}. However, the dips were not present in the {\it XMM-Newton} observation conducted in 2011, where the source identified in a low-flux state. The {\it Swift} and {\it Chandra} spectra showed the spectral variability similar to that of other ULXs, and spectral parameters such as absorption column density and normalization of a two-component model varied in these observations \citep{Pin15}. The RGS spectral analysis of NGC~55~ULX1 showed blueshifted emission and absorption lines, which indicate the presence of a strong wind \citep{Pin17}. The observed temporal and spectral behaviour is in agreement with the scenario of clumpy wind of obscuring material entering the line of sight \citep{Sto04, Mid15a, Pin17}.  

The earlier X-ray observations of NGC~55~ULX1 revealed the presence of outflows and significant X-ray flux dips from the source \citep{Sto04, Pin15, Pin17}. In this work, we revisit this source using {\it Swift}, {\it XMM-Newton} and {\it NuSTAR} observations conducted after the year 2013 to investigate its spectral and temporal characteristics. Section \ref{sec:obs} describes the observations used in this work and the data reduction techniques. The results from the temporal and spectral analysis are presented in \S \ref{sec:analysis}. The main results are discussed in \S \ref{sec:discu}.

\section{Observations and Data Reduction}
\label{sec:obs}
We used the X-ray data of NGC~55~ULX1 from the {\it Neil Gehrels Swift Observatory} \citep[{\it Swift};][]{Geh04}, {\it XMM-Newton} \citep{Jan01} and {\it NuSTAR} \citep{Har13} observatories during the period 2013 September -- 2021 May. The observations used in the work are listed in Table \ref{obslog}.

\begin{table*}
\centering
\setlength{\tabcolsep}{8.0pt}
	\caption{Observation log. (1) Mission; (2) data; (3) observation ID; (4) date/period of observation; (5) exposure time; (6) the background-subtracted source count rates. The three count rate values in the {\it XMM-Newton} data are from PN, MOS1 and MOS2, respectively. In the case of {\it NuSTAR} observations, the two count rates are from FPMA and FPMB, respectively. The {\it Swift} and {\it XMM-Newton} count rates are in the 0.3--10 keV energy band, while the {\it NuSTAR} count rates are in the 3--20 keV energy range. The prefix A, B, C, D, E and F on {\it Swift} denotes 000821200, 000334680, 000923050, 0008895, 000800020 and 000326190, respectively.} 
 	\begin{tabular}{@{}ccccccc@{}}
	\hline
	\hline
Mission & Data & Obs ID & Date & Exposure & CR \\
 & & & & (ks) & (counts s$^{-1}$) \\
\hline
{\it Swift} & & A01-A05 & 2013 Sep - 2014 Jan & 0.2-3.5 & \\
 & & B01,B03-B06,B09 & 2014 Oct &  0.5-3.7 & \\
 & & A06-A07 & 2015 Apr & 2.5-3.7 & \\
 & & C01, C03-C45 & 2016 Sep & 0.4-1.1 & \\
 & & C45-C50 & 2016 Oct & 0.04-1 & \\
 & & B10-B18, B21-B22, B24-B26, B29-B33, B35 & 2016 Nov & 0.1-0.7 & \\
 & & B36-B38 & 2016 Dec & 0.6 & \\
 & & D2001-D3001,D3002 & 2019 Nov - Dec & 0.7-1.1 & \\
 & & E01 & 2020 May 01 & 1.3 & \\
 & & F21-F24 & 2020 May  & 1.5 & \\
 & & E02-E20 & 2020 May - 2021 Jan & 0.7-1.6 & \\
 & & F25-F34 & 2021 Jan - May & 0.5-1.6 & \\

& {\it Swift1} & & & 51.2 & $3.1 \times 10^{-2}$ \\
& {\it Swift2} & & & 43.5 & $4.7 \times 10^{-2}$ \\ 
& {\it Swift3} & & & 23.7 & $7.7 \times 10^{-2}$ \\

{\it XMM-Newton} & XMM1 & 0824570101 & 2018 Nov 17 & 139.8 & 0.59/0.16/0.16\\ 
 & XMM2 & 0852610101 & 2019 Nov 27 & 11 & 1.25/0.34/0.38\\
 & XMM3 & 0852610201 & 2019 Dec 27 & 8 & 1.23/0.34/0.35\\
 & XMM4 & 0852610301 & 2020 May 11 & 9 & 0.48/0.14/0.15\\
 & XMM5 & 0852610401 & 2020 May 19 & 8 & 0.22/0.15/0.25\\
 & XMM6 & 0864810101 & 2020 May 24 & 132.8 & 0.82/0.20/0.21\\
{\it NuSTAR} & N1 & 50510001002 & 2019 Nov 26 & 52.8 & $5.8 \times 10^{-3}$/$5.2 \times 10^{-3}$ \\ 
 & N2 & 50510003002 & 2020 May 11 & 48.8 & $7.7 \times 10^{-4}$/$1.7 \times 10^{-3}$\\

\hline
\end{tabular} 
\label{obslog}
\end{table*}

\subsection{{\textbf{\it Swift}}}
We have analysed 119 {\it Swift} X-ray Telescope (XRT) observations of NGC 55 conducted during the period 2013 September--2021 May. The XRT data were processed, filtered and screened with the {\sc xrtpipeline} (version: 0.13.5) using the standard criteria. We used the photon-counting (PC) mode data from all the observations. The source and background events were extracted from a circular region with a radius of 47'' and 94'' using the standard grade filtering of 0--12. We used the standard calibration database (CALDB) spectral redistribution matrices (version: 20200724), while the ancillary response files were generated with {\sc xrtmkarf}, which accounts for different extraction regions, vignetting and PSF corrections. Among 119 observations, two observations (Obs ID 00092305048 and 00032619028) have no PC data, and for the observation 00092305049, the source is undetectable in the PC data. Thus, we have not included these observations in the further analysis.

\begin{figure*}
\begin{center}

\includegraphics[width=17cm,angle=0]{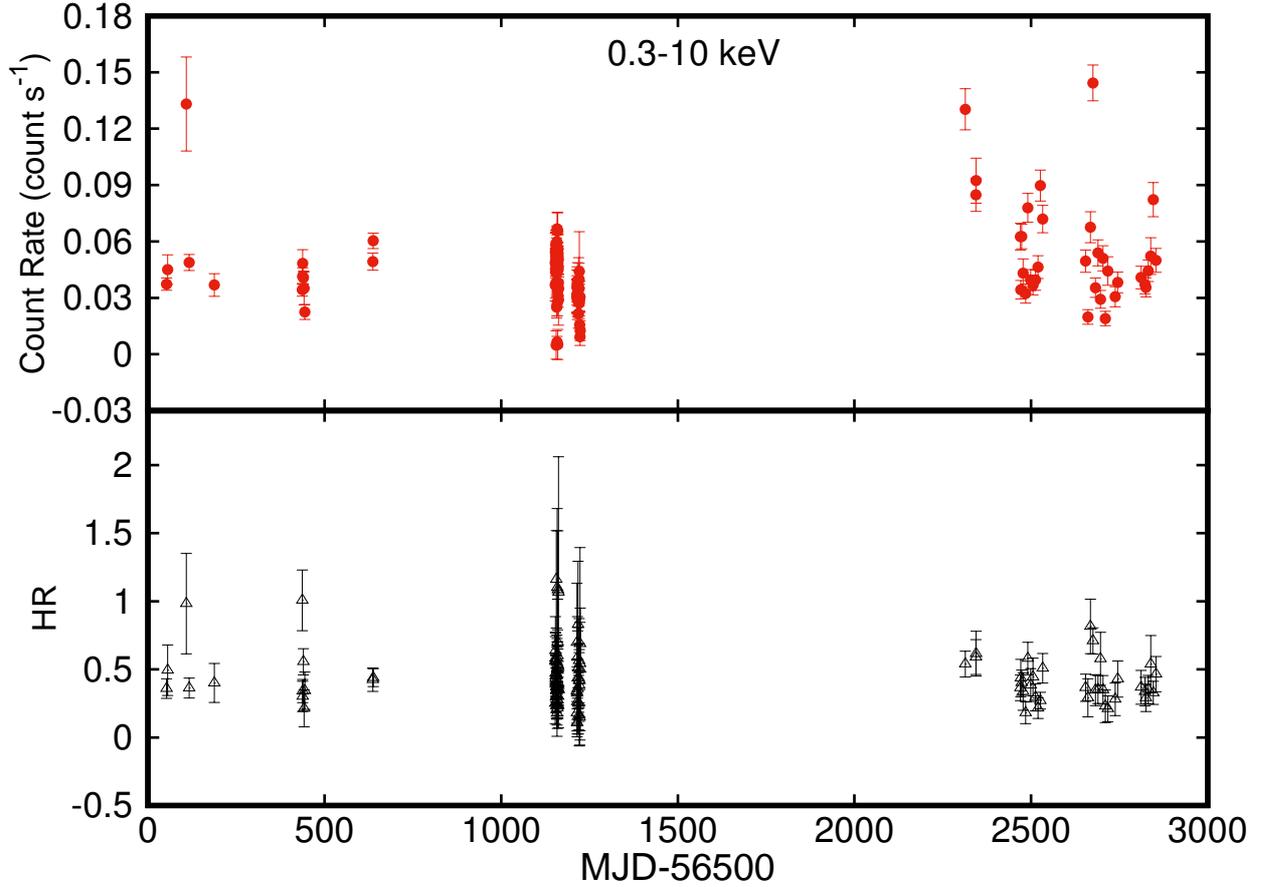}
\caption{The 0.3--10 keV {\it Swift}-XRT light curve (top panel) and hardness ratio (bottom panel) of NGC~55~ULX1 over a period 2013 September 18 to 2021 May 31. Each data point corresponds to individual observation. The hardness ratio is derived from 0.3--1.5 and 1.5--10 keV energy bands.}
\label{swift-lc}
\end{center}
\end{figure*}

\subsection{{\textbf{\it XMM-Newton}}}

We obtained six publicly available {\it XMM-Newton} observations from the High Energy Astrophysics Science Archive Research Center (HEASARC) database \footnote{\url{https://heasarc.gsfc.nasa.gov}} conducted during 2018--2020. We used the XMM-Newton European Photon Imaging Camera (EPIC) PN \citep{Str01} and metal oxide semiconductor \citep[MOS;][]{Tur01} instrument data for the analysis. The EPIC PN and MOS data were reduced using the standard tools ({\sc epchain} and {\sc emchain}, respectively) of XMM-Newton Science Analysis Software ({\sc sas}; version 18.0). We removed the particle flaring background by extracting the full-field background light curve from the EPIC camera in the 10--12 keV energy band and created good time interval files for each observation. We filtered the PN data using the flag expression \#XMMEA\_EP and PATTERN $\leq 4$, and used FLAG==0, PATTERN $\leq 12$ and \#XMMEA\_EM expression for MOS data. The source spectrum is extracted from a circular region of radius 40'' centred at the source position \citep[R.A.=00:15:28.9, decl.=-39:13:19, equinox J2000.0;][] {Gla09}. For the background spectrum, a source-free region in the same CCD, close to the source, is used and the radius is the same as that of the source region. The source spectrum, background spectrum, ancillary response file (ARF) and redistribution matrix file (RMF) extracted using the {\sc sas} task {\sc especget}. We note that the source resides on the chip-gap in the XMM5 observation and the X-ray events in the chip-gap have poor spectral calibration. In order to mitigate these events, we used the selection criterion ``FLAG==0'' \citep[see][for more details]{Dew10, Jit18}. This method significantly reduces the source flux, however, the flux loss can be corrected by the {\sc sas} task {\sc arfgen}\footnote{\url{https://xmm-tools.cosmos.esa.int/external/xmm\_user\_support/documentation/sas\_usg/USG/}}. We used a minimum count of 100 per bin for the observations XMM1 and XMM6, while for the rest of the observations we used a minimum count of 20 per bin. We extracted the source and background light curves from the same regions used for the spectral products. The extracted light curve is corrected using the {\sc sas} tool {\sc epiclccorr}.

\subsection{{\textbf{\it NuSTAR}}}

{\it NuSTAR} observed NGC 55 on four occasions (Obs ID: 50510001002, 50510002002, 50510003002 and 50510004002) during the period 2019--2020 and NGC 55 ULX1 covered in two observations 50510001002 and 50510003002. Thus, we used only these observations for the rest of the analysis. We reprocessed the data with {\sc nupipeline} task available in the {\it NuSTAR} Data Analysis Software ({\sc nustardas}; version 2.0.0). We applied the standard filtering criteria to obtain the cleaned event files and used the {\it NuSTAR} CALDB version 20200813. The high-level science products such as spectra and light curves were extracted using {\sc heasoft} (version 6.28) tools. The source and background spectra were extracted for both focal plane modules (FPMA and FPMB) separately from circular regions of radius 40 arcsec. We grouped the spectra using {\sc grppha} with a minimum of 20 counts per bin. Since the source is weak in the hard energies and the background dominates above 20 keV energy, we restrict the {\it NuSTAR} spectral analysis in the 3--20 keV energy band. We fitted both FPMA and FPMB spectra jointly.

\section{Analysis and Results}
\label{sec:analysis}
\subsection{Temporal Analysis}
\label{sec:temporal}

\begin{figure}
\begin{center}

\includegraphics[width=8.5cm,angle=0]{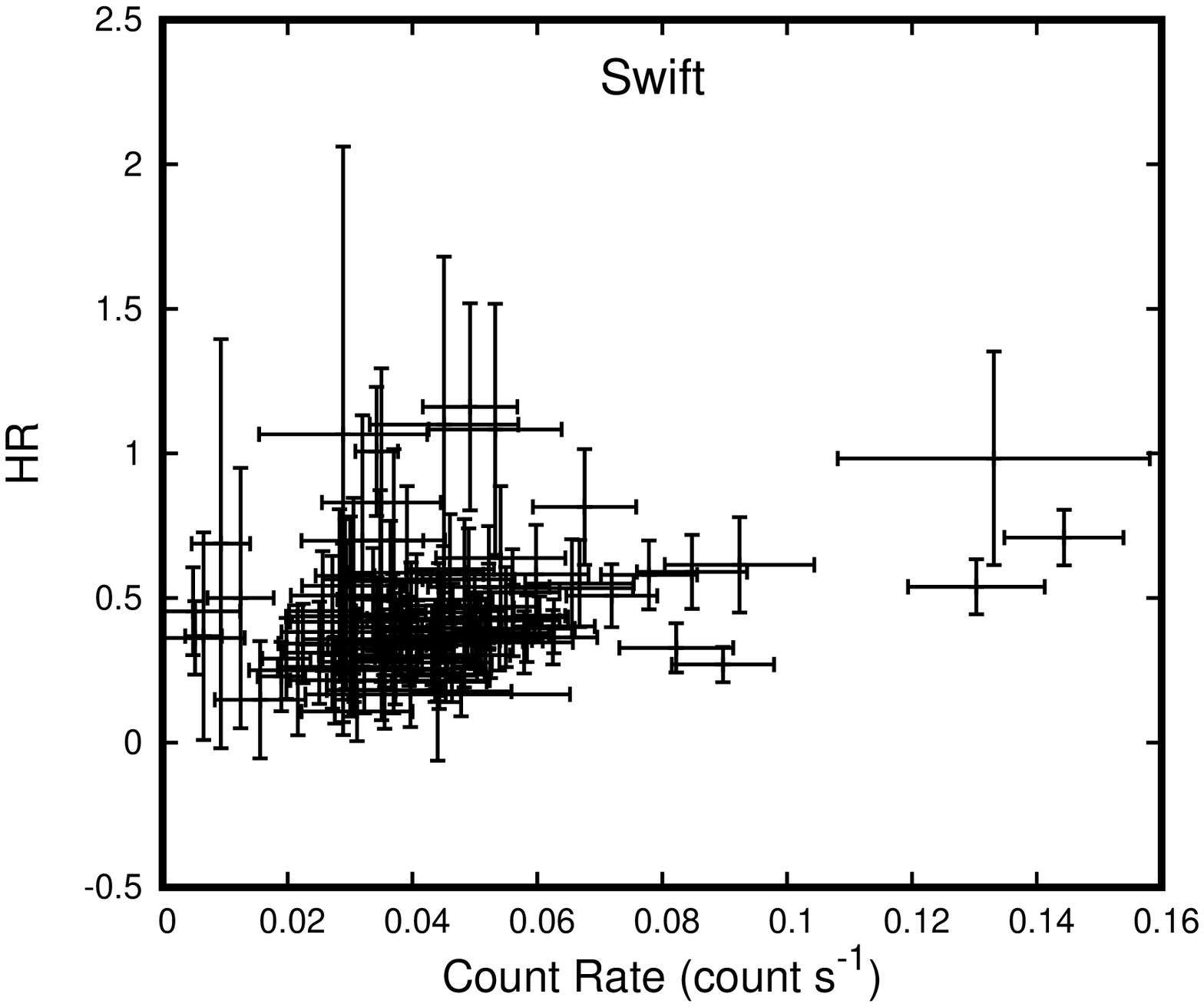}
\includegraphics[width=8.5cm,angle=0]{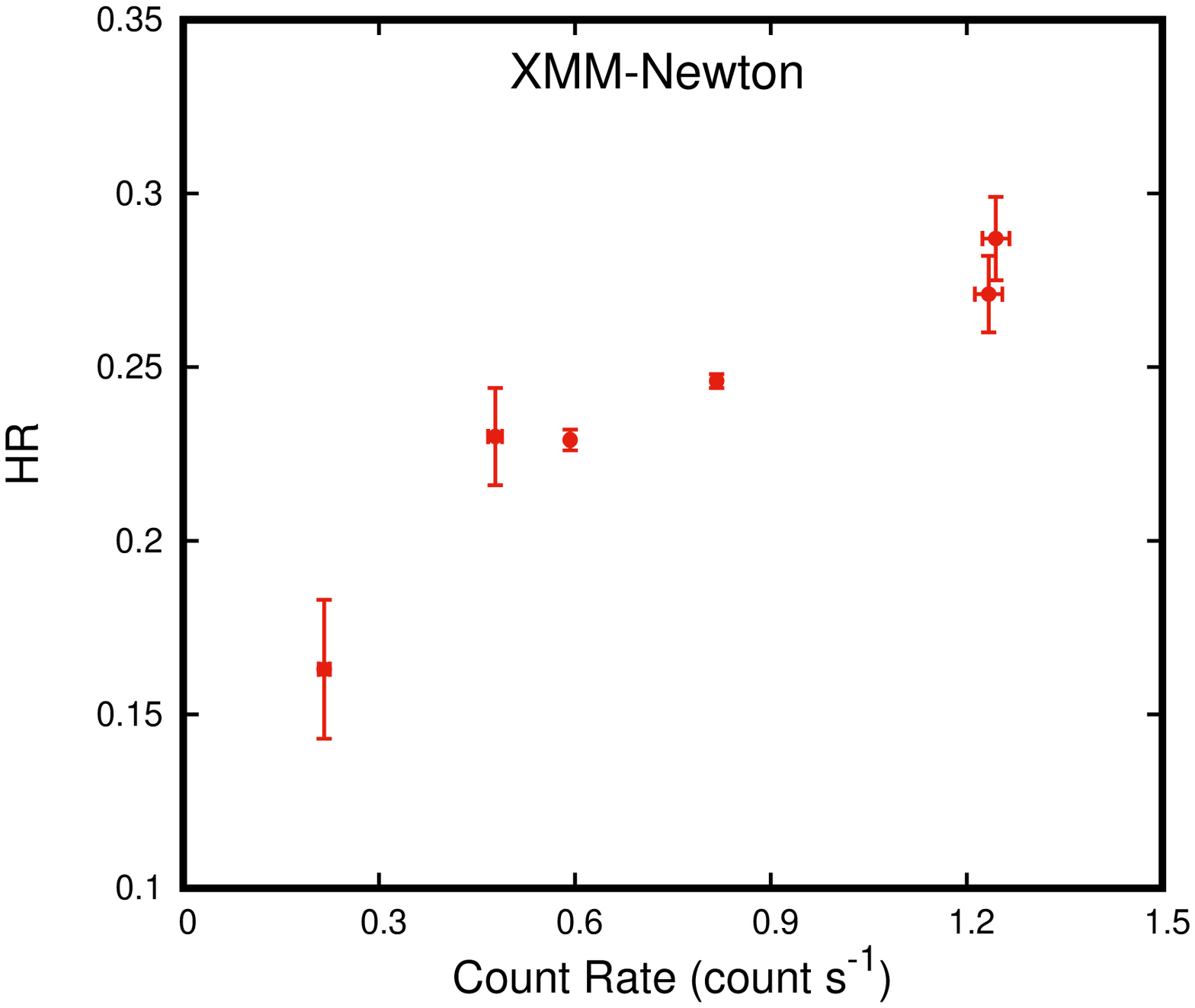}
\caption{The Hardness-Intensity diagram of NGC~55~ULX1 using {\it Swift} XRT (top) and {\it XMM-Newton} (bottom) observations. For {\it XMM-Newton} observations, we used the EPIC PN data.}

\label{HID}
\end{center}
\end{figure}

To explore the long-term evolution of NGC 55 ULX1, we reprocessed the 116 short {\it Swift} XRT observations and plotted the long-term light curve, which is shown in the top panel of Figure \ref{swift-lc}. The hardness ratio (HR) is defined as the ratio of count rate from 1.5--10 keV and 0.3--1.5 keV energy bands, which is plotted in the lower panel of Figure \ref{swift-lc} for individual observation. From Figure \ref{swift-lc}, it is clear that the source is variable by a factor of $\sim 5-6$ in intensity. We showed the variation of HR with the intensity in the top panel of Figure \ref{HID}. The HR marginally varies in these short observations but with large uncertainties. In {\it XMM-Newton} observations, we do observe a change in the intensity by a factor $\sim 6$ similar to {\it Swift} XRT observations, and the HR varies from 0.16 to 0.29 as the intensity increases (see the bottom panel of Figure \ref{HID}). 

\begin{figure}

 \includegraphics[width=8.1cm,angle=0]{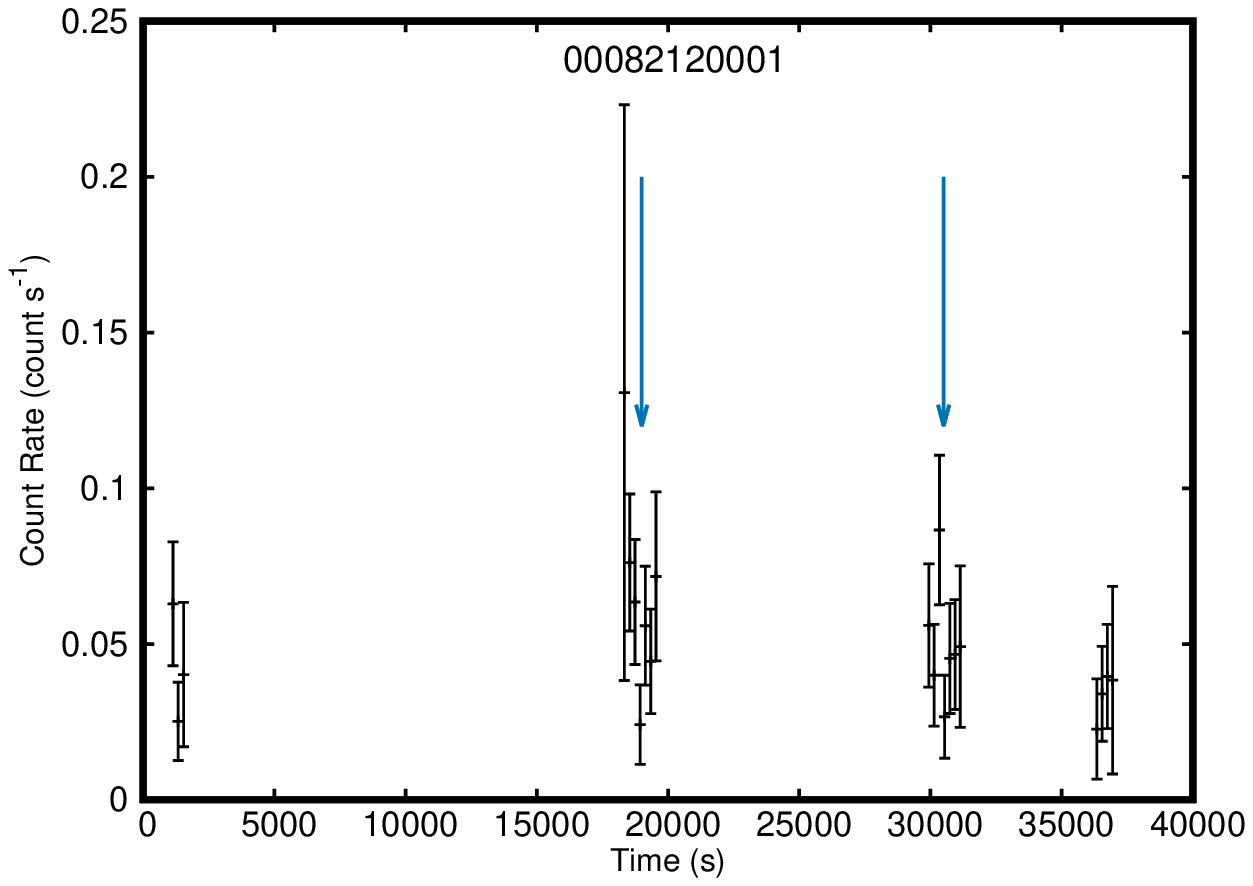}
 \includegraphics[width=8.2cm,angle=0]{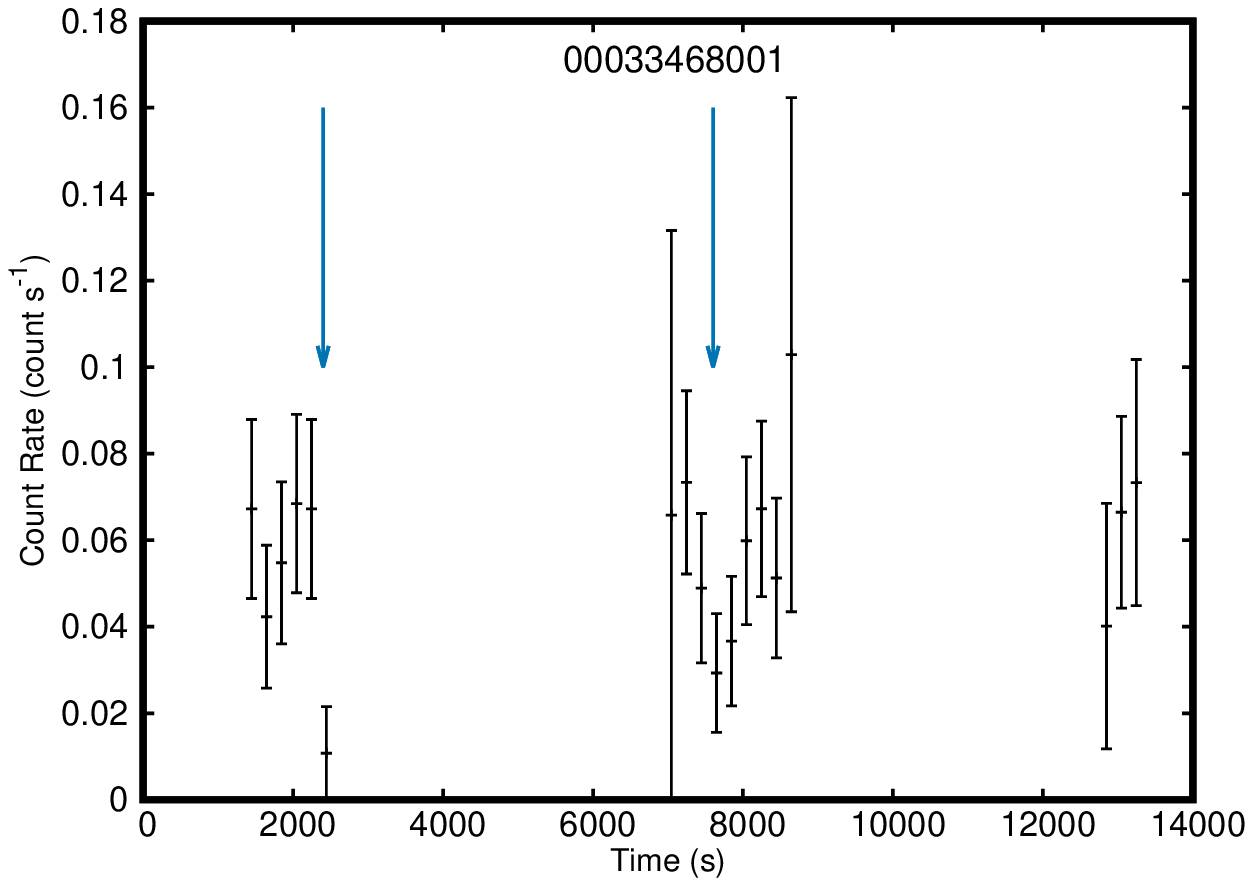}
 \includegraphics[width=8.3cm,angle=0]{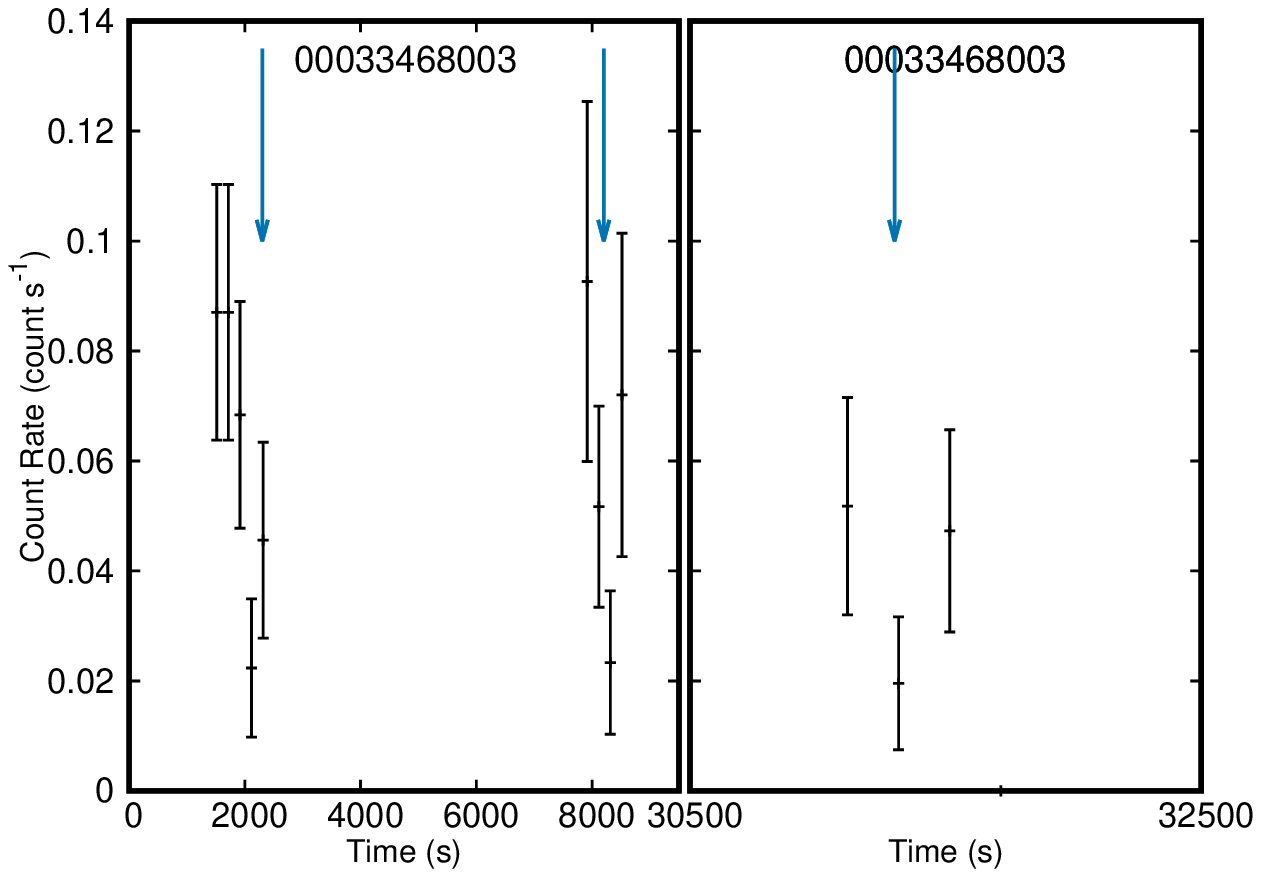}
 \includegraphics[width=8.30cm,angle=0]{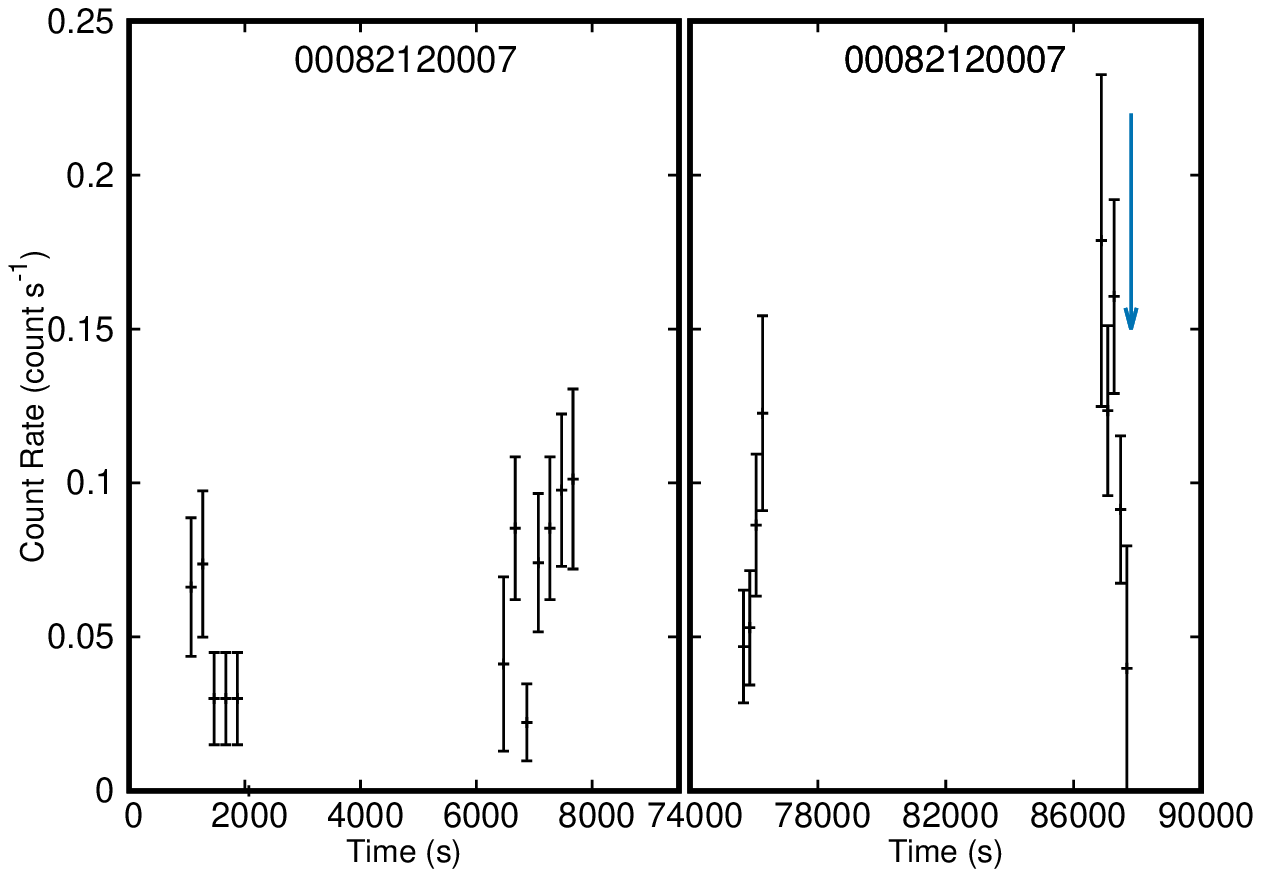}

\caption{The 0.3--10 keV light curves of NGC 55 ULX1 from {\it Swift} observations. The light curve is binned with 200 s. The vertical arrows represent the probable dips from the source.}

\label{lc_swift}
\end{figure}

\begin{figure*}

 \includegraphics[width=18.0cm,angle=0]{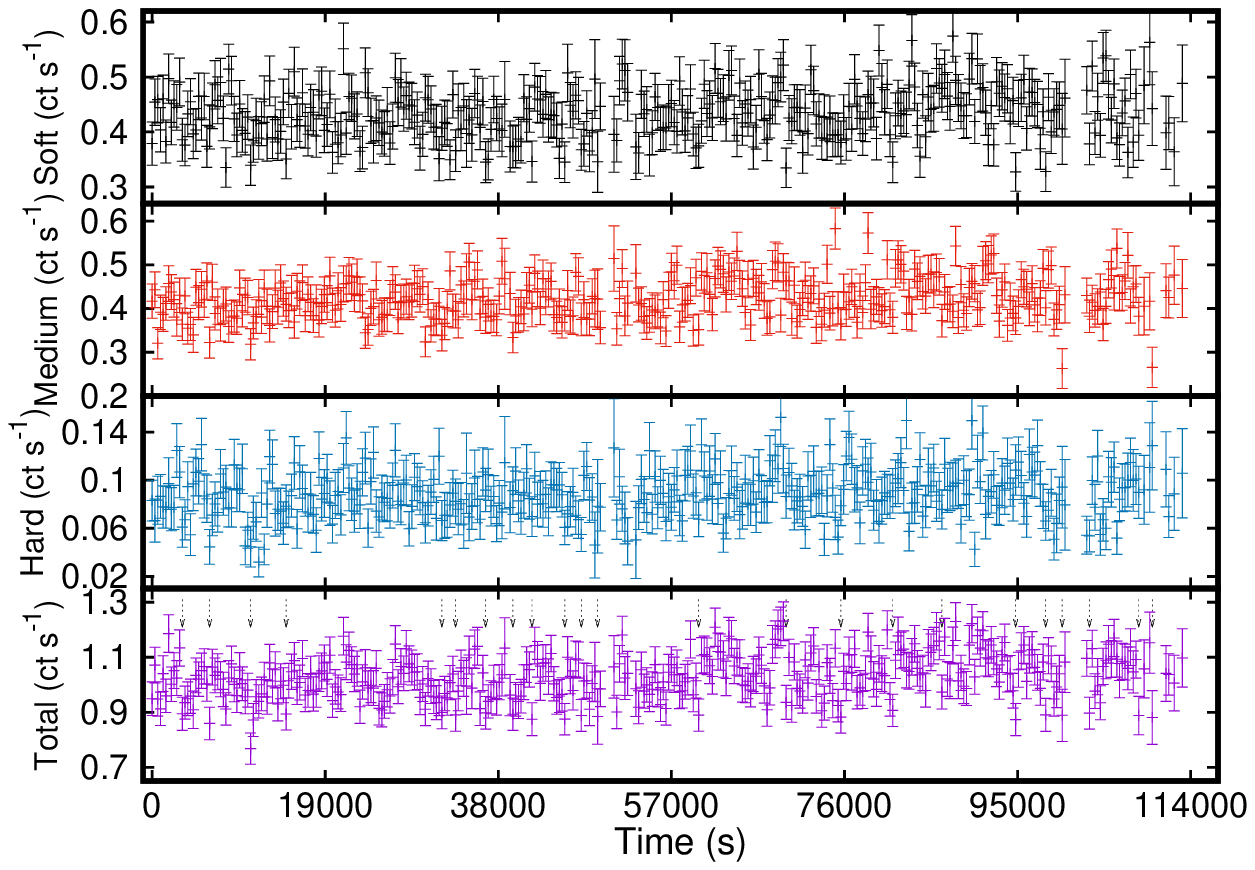}

\caption{The light curve of NGC 55 ULX1 from XMM1 observation binned with 300 s in different energy bands. From top to bottom: 0.5--1.0 keV (soft), 1.0--2.0 keV (medium), 2.0--4.5 keV (hard) and 0.3--10 keV (total). These light curves are based on the combined data from the MOS1, MOS2 and PN cameras. The dashed vertical lines in the 0.3--10 keV light curve represent the possible dips from the source. The RMS variability is $<3\%$ (at $3\sigma$ level), $(5.1 \pm 0.9)\%$, $<11.2\%$ (at $3\sigma$) and $(4.4 \pm 0.5)\%$ in the soft, medium, hard and total energy bands, respectively.}

\label{lc_xmm1}
\end{figure*}

\begin{figure*}

 \includegraphics[width=18.0cm,angle=0]{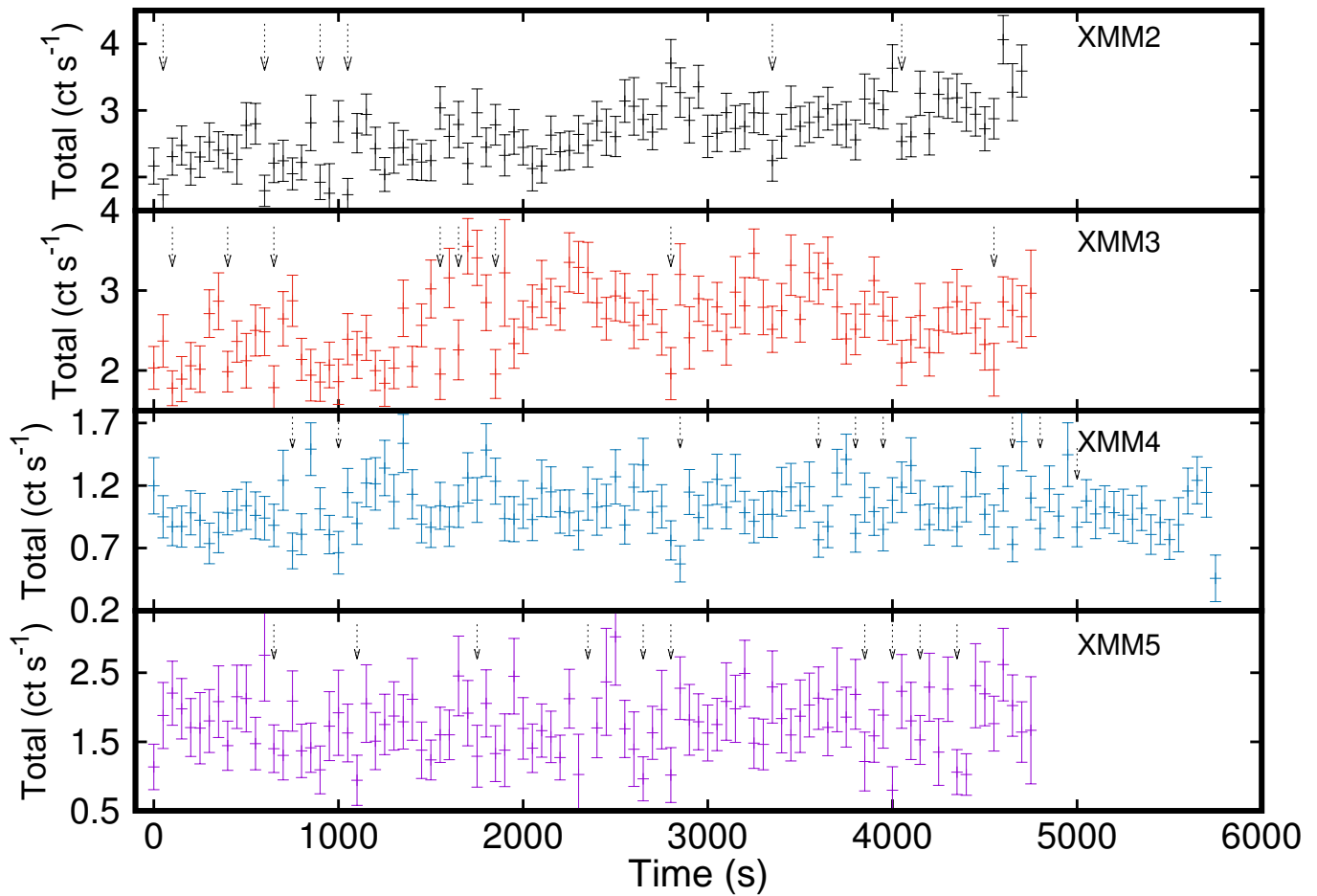}

\caption{The 0.3--10 keV light curve of NGC 55 ULX1 from XMM2 (top), XMM3 (upper middle), XMM4 (lower middle) and XMM5 (bottom) observations. The bin size is 50 s. These light curves are based on the combined data from the MOS1, MOS2 and PN cameras. The dashed vertical lines represent the possible dips from the source. The RMS variability is $(13.9 \pm 1.8)\%$, $(11.4 \pm 1.8)\%$, $<10.0\%$ (at $3\sigma$) and $<19.9\%$ (at $3\sigma$) in the XMM2-5 observations, respectively.}

\label{lc_xmm2345}
\end{figure*}

\begin{figure*}

 \includegraphics[width=18.0cm,angle=0]{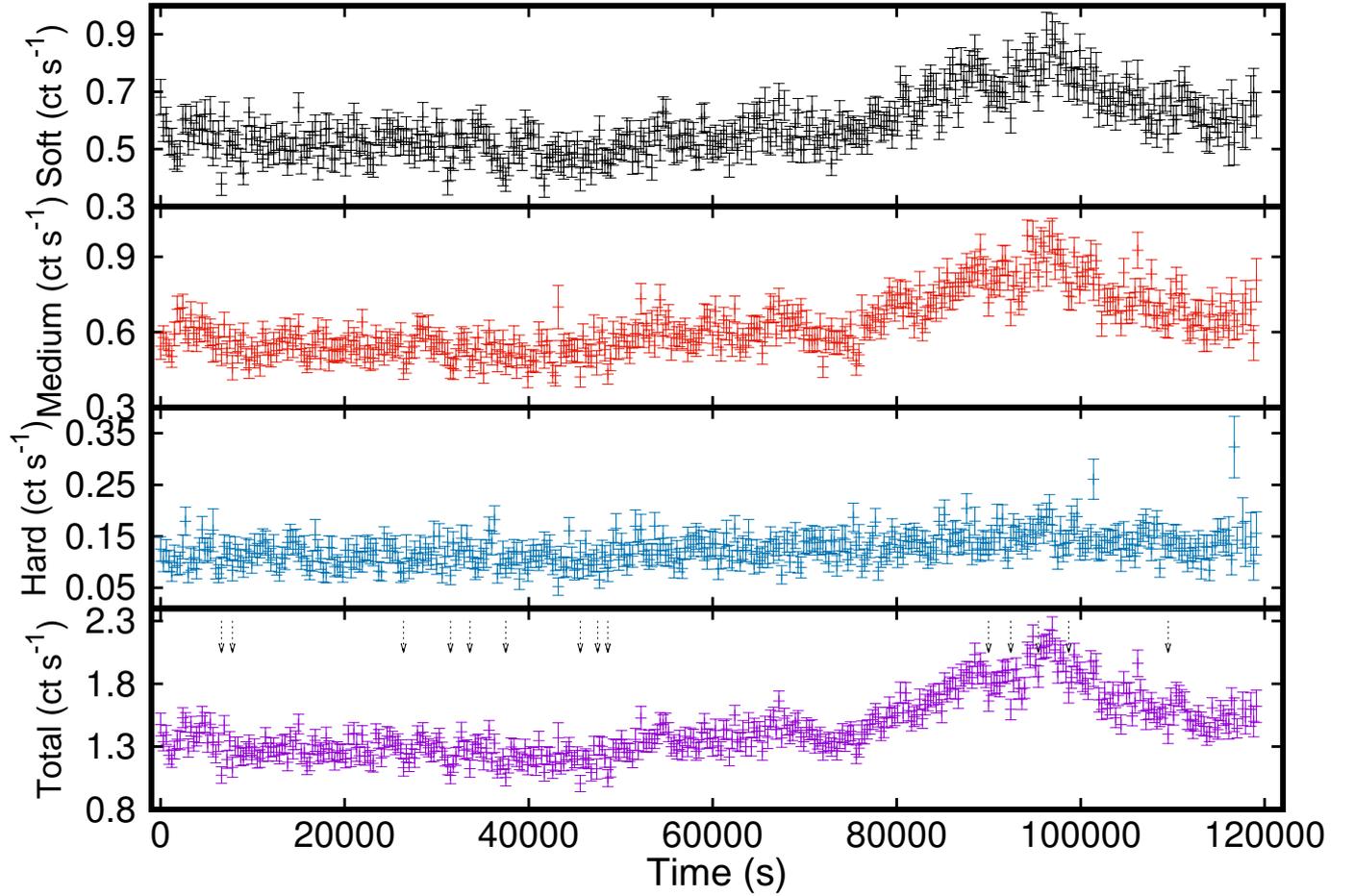}

\caption{The light curve of NGC 55 ULX1 from XMM6 observation binned with 300 s in different energy bands. From top to bottom: 0.5--1.0 keV (soft), 1.0--2.0 keV (medium), 2.0--4.5 keV (hard) and 0.3--10 keV (total). These light curves are based on the combined data from the MOS1, MOS2 and PN cameras. The dashed vertical lines in the 0.3--10 keV light curve represent the possible dips from the source. The RMS variability is $(15.1 \pm 0.7)\%$, $(16.9 \pm 0.7)\%$, $(12.7 \pm 1.4)\%$ and $(15.5 \pm 0.6)\%$ in the soft, medium, hard and total energy bands, respectively.} 

\label{lc_xmm6}
\end{figure*}

NGC 55 ULX1 exhibited significant X-ray dipping behaviour in the {\it XMM-Newton} observations and individual dips lasted for typically 100--300\,s \citep[see Figure 2 and 4 of][]{Sto04}. To search the X-ray dips, we first checked the {\it Swift} observations with an exposure $> 2$\,ks. We observed marginal evidence of X-ray dipping events in four (Obs ID 00082120001, 00033468001, 00033468003 and 00082120007) out of seven {\it Swift} observations with an exposure $> 2$\,ks. For these observations, the average count rate between $\sim 0.05-0.07$ counts s$^{-1}$ in the 0.3--10 keV light curve and the dips are at the level of $\sim 0.01-0.02$ counts s$^{-1}$ (see Figure \ref{lc_swift}). These probable dips are similar to that observed in the {\it Swift} observations conducted in 2013 \citep[See Figure 4 of][]{Pin15}. To quantify the variability in the {\it Swift} observations, we computed the RMS variability of the light curve using the tool {\sc lcstats}. Given the quality of the data, we obtained RMS upper limits in these observations except one (Obs ID 00082120007). The computed RMS is $<65\%$ (Obs ID 00082120001), $<67\%$ (Obs ID 00033468001), $<34\%$ (Obs ID 00033468003) and $(48 \pm 11)\%$ (Obs ID 00082120007). The obtained upper limits are at $3\sigma$ level. We further explored the dipping features in the {\it XMM-Newton} light curves by combining the data from the MOS1, MOS2 and PN cameras as done in \citet{Sto04}. We extracted the light curves in soft (0.5--1 keV), medium (1--2 keV), hard (2--4.5 keV) and total (0.3--10 keV) energy bands using different time bins of 50-300\,s depending on the length of the observations. In these {\it XMM-Newton} observations, we did not find significant X-ray flux dips of that quality observed in the 2001 {\it XMM-Newton} observations. However, we found marginal evidence for dips in the 0.3--10 keV light curve and they are shown in Figure \ref{lc_xmm1}, \ref{lc_xmm2345} and \ref{lc_xmm6}. The average count rate (in the 0.3--10 keV band) is in the range of $\sim 1.0-2.7$ counts s$^{-1}$, and during the probable dipping events, the count rate is dropped by $\sim 5-73$\% from the average values. The depth of the dips is slightly greater in the hard band (2--4.5 keV) compared to other energy bands, and the intensity dropped to $\sim 58-99$\% of the average value in the hard band. In addition, we extracted the light curve with different time bins say 200 and 400s from XMM1 and XMM6 observations. We observed marginal evidence of dips in these light curves as seen in the 300s binned light curve. In addition, we computed the RMS variability of {\it XMM-Newton} light curves using {\sc lcstats} and the estimated variability is $\sim 4-16\%$ in the different energy bands (see Figure \ref{lc_xmm1}, \ref{lc_xmm2345} and \ref{lc_xmm6}).

To search for pulsating signals from the source, we used {\it XMM-Newton} and {\it NuSTAR} observations. We first applied the barycentric corrections to {\it XMM-Newton} and {\it NuSTAR} event files using {\sc sas} task {\tt barycen} and {\sc ftool} {\tt barycorr} tool, respectively. We extracted the power spectra from the {\it XMM-Newton} and {\it NuSTAR} observations using the {\sc ftool} {\tt powspec} in the 0.001--10 Hz frequency range. However, we did not find any signals in these power spectra. In addition, we used the {\tt HENDRICS}\footnote{\url{https://hendrics.stingray.science/en/latest/}} software package \citep{2015ascl.soft02021B, 2019ApJ...881...39H} to search the signals from {\it NuSTAR} observations in the 0.001--10 Hz frequency range. This software properly treats the data gaps in the observation and dead time in the {\it NuSTAR} detectors. This analysis also did not provide any strong pulsating signals. The $5\sigma$ upper limit on the RMS variability is $<40\%$ in the N1 observation, while for the N2 data, we could not derive a meaningful result because of the poor data quality.

\subsection{Spectral Analysis}
\label{sec:spectral}

We used the 116 {\it Swift} XRT, six {\it XMM-Newton} and two {\it NuSTAR} observations for the spectral analysis. In the case of {\it Swift} XRT, the individual observation is short and not enough spectral counts are available for modelling. Thus, to increase the signal to noise and perform the spectral modelling with simple models, we decided to add the {\it Swift} XRT data. For this, we defined three intensity ranges: 0--0.04 ($Swift$1), 0.04--0.055 ($Swift$2) and 0.055--0.15 ($Swift$3) counts s$^{-1}$, to get roughly equal numbers of total counts (within a factor of 2) in the added spectrum. We added the individual spectrum in the each count rate range using the tool {\sc addascaspec}. We performed the spectral modelling in the 0.3--10 keV energy band using {\sc xspec} version 12.11.1 \citep{Arn96}. Two absorption components ({\tt tbabs} in {\sc xspec}) were considered in all spectral modelling. The first absorption component is fixed at the Galactic absorption value, $7.6 \times 10^{20}\rm~cm^{-2}$ \citep{HI416}, while the second one is considered as a free parameter, which describes the local absorption.    



We first used the two-component model: multi-colour disc blackbody (MCD; {\tt diskbb}) component plus power law (PL) component for the co-added {\it Swift} XRT and {\it XMM-Newton} spectra. The best-fit spectral parameters of this model are given in Table \ref{diskbb_PL_params}. All quoted errors were computed at a 90\% confidence level. This model provides a satisfactory fit for all the spectra except XMM1 and XMM6 observations. In XMM1 and XMM6, the fit provided a reduced $\chi^2$ ($\chi{^2}_{r} = \chi^2$/degrees of freedom (d.o.f)) of $\sim 1.9$ and $\sim 2.1$, respectively. In the case of satisfactory fit, the inner disc temperature is in the range of 0.07--0.13 keV and the power law index of 3.08--4.38. We then attempted a more complex model, namely a combination of an MCD and a Comptonization component with a roll-over at high energy ({\tt comptt}), that is typically used to describe the ULXs spectra \citep{Sto06, Gla09}. Since the {\it Swift} XRT spectra are of low quality, we applied the disc plus Comptonization model only to the high quality {\it XMM-Newton} spectra. In the model fit, we set the seed photon temperature ($\rm T_{0}$) equal to the inner disc temperature ($\rm T_{in}$). The model provides an improved fit compared to the MCD plus PL model in all spectra, and the $\Delta\chi{^2}$ is 6.6--466.2 with the loss of one additional d.o.f. In addition, the best-fit values of inner disc temperature and the plasma electron temperature are in the range of 0.14--0.19 keV and 0.5--1.2 keV, respectively, while the optical depth is not well constrained in some of the {\it XMM-Newton} observations. 

\begin{table*}
\setlength{\tabcolsep}{18.0pt}
	\caption{The best-fit X-ray spectral parameters for NGC~55~ULX1 from the doubly absorbed MCD plus PL model. (1) Observation; (2) neutral hydrogen column density from {\tt tbabs} model; (3) inner disc temperature; (4) MCD normalization; (5) the PL index; (6) normalization of the PL; (7) $\chi^2$ statistics and degrees of freedom. $^{a}$ The reduced $\chi^2$ is $> 2$ for XMM6.}
 	\begin{tabular}{@{}ccccccc@{}}
	\hline
	\hline
Obs & $N_{\rm H}$ & $\rm kT_{in}$ &$\rm N_{diskbb}$ & $\Gamma$ &$\rm N_{PL}$ & $\rm \chi^2/ d.o.f$ \\
& ($10^{22}~\rm cm^{-2}$) & (keV) & ($10^{4}$) & & ($10^{-3}$) & \\
\hline

$Swift$1 & $0.61^{+0.28}_{-0.25}$ & $0.11^{+0.05}_{-0.02}$ & $3.74^{+156}_{-3.65}$ & $3.02^{+0.75}_{-0.84}$ & $0.79^{+0.70}_{-0.42}$ & 48.4/62 \\ 
$Swift$2 & $1.13^{+0.22}_{-0.23}$ & $0.07^{+0.01}_{-0.01}$ & $3400^{+39300}_{-3210}$ & $4.38^{+0.38}_{-0.32}$ & $3.59^{+1.57}_{-1.19}$ & 73.1 75 \\ 
$Swift$3 & $0.66^{+0.29}_{-0.29}$ & $0.10^{+0.05}_{-0.02}$ & $15.0^{+544}_{-15.0}$ & $3.57^{+0.38}_{-0.13}$ & $2.81^{+1.59}_{-0.93}$ & 67.5 70 \\ 

XMM1 & $0.44^{+0.02}_{-0.02}$ & $0.12^{+0.00}_{-0.00}$ & $0.52^{+0.15}_{-0.17}$ & $3.64^{+0.03}_{-0.04}$ & $1.04^{+0.03}_{-0.05}$ & 1193.1/599 \\ 
XMM2 & $0.45^{+0.08}_{-0.06}$ & $0.12^{+0.02}_{-0.02}$ & $0.91^{+3.60}_{-0.71}$ & $3.71^{+0.14}_{-0.13}$ & $3.20^{+0.49}_{-0.39}$ & 437.7/368 \\ 
XMM3 & $0.47^{+0.10}_{-0.07}$ & $0.12^{+0.02}_{-0.01}$ & $2.08^{+10.1}_{-1.68}$ & $3.85^{+0.16}_{-0.14}$ & $3.38^{+0.62}_{-0.46}$ & 318.7/295 \\
XMM4 & $0.51^{+0.16}_{-0.11}$ & $0.11^{+0.02}_{-0.02}$ & $1.45^{+13.8}_{-1.24}$ & $3.72^{+0.29}_{-0.25}$ & $1.13^{+0.38}_{-0.43}$ & 215.3/169 \\
XMM5 & $0.42^{+0.14}_{-0.08}$ & $0.13^{+0.04}_{-0.03}$ & $0.48^{+7.41}_{-0.44}$ & $3.99^{+0.30}_{-0.28}$ & $1.98^{+0.61}_{-0.47}$ & 151.3/127 \\
XMM6 & 0.47 & 0.12 & 0.86 & 3.81 & 1.76 & 1446.0/687$^{a}$ \\

\hline
\end{tabular} 
\label{diskbb_PL_params}
\end{table*}

\begin{figure*}
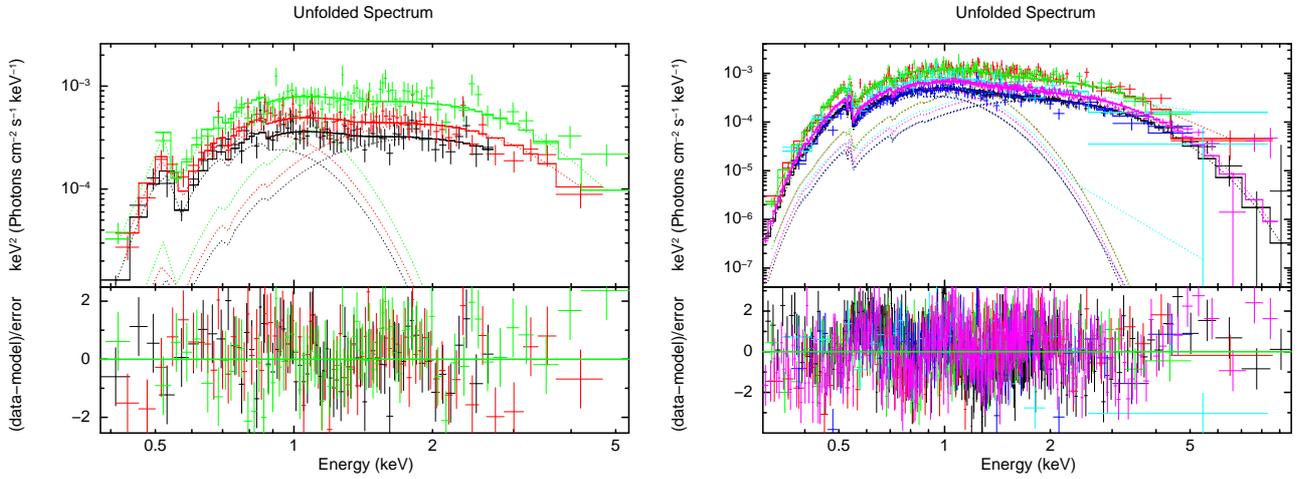


 \includegraphics[width=6.3cm,angle=-90]{3swift_spectra_2tbabs_bbody_diskbb_eeufspec.ps}
 \includegraphics[width=6.3cm,angle=-90]{6xmm_spectra_2tbabs_bbody_diskbb_eeufspec.ps}

\caption{Three {\it Swift}-XRT (left) and six {\it XMM-Newton} (right) spectra of NGC 55 ULX1 in the 0.3--10 keV energy band. In {\it Swift}, three spectra are from different intensity levels: 0--0.04 (black), 0.04--0.055 (red) and 0.055--0.15 (green) count s$^{-1}$. In {\it XMM-Newton} observations, the black, red, green, blue, cyan and pink data points represent XMM1-6, respectively. We showed the EPIC PN spectrum only for clarity. The spectra are fitted with doubly absorbed blackbody and disc blackbody components.}
\label{3swift_6xmm_spec}
\end{figure*}

We then used another two-component model consisting of a blackbody ({\tt bbody}) component and a disc ({\tt diskbb}) component, which can be considered as a proxy for more complex models \citep[see, e.g.,][]{Wal14, Pin15, Pin17}. Since it is a simple model, we applied the model to the co-added {\it Swift} XRT and {\it XMM-Newton} spectra (see Figure \ref{3swift_6xmm_spec}). This model improved the spectral fit in all spectra compared to the {\tt diskbb} plus PL model and the $\Delta\chi{^2}$ is 2.5--577.7 with no change in the d.o.f for the co-added {\it Swift} XRT and {\it XMM-Newton} spectra. Compared to the complex model (disc plus Comptonization model), the {\tt bbody} plus {\tt diskbb} model provides an improved fit (four out of six cases) or slightly worse fit in the case of {\it XMM-Newton} spectra. We derived the unabsorbed flux using the {\tt cflux} model in {\sc xspec} and computed the unabsorbed luminosity by assuming the source distance of 1.78 Mpc \citep{Kar03}. The luminosity and best-fit parameters are given in Table \ref{bbody_diskbb_params}. In XMM6, the source exhibits intensity variation within the observation. Thus, we performed the flux-resolved spectral analysis by choosing two intensity levels with count rates below and above the average count rate values for PN ($\sim$ 0.92 counts s$^{-1}$), MOS1 ($\sim$ 0.25 counts s$^{-1}$)and MOS2 ($\sim$ 0.27 counts s$^{-1}$). We observe a marginal decrease in the absorption, while the blackbody temperature increases slightly from low to high-intensity levels. The inner disc temperature and the unabsorbed luminosity show an increasing trend, however, the values are consistent within the uncertainty. 

\begin{table*}
\setlength{\tabcolsep}{10.0pt}
	\caption{The X-ray spectral parameters for NGC~55~ULX1 from the doubly absorbed blackbody plus disc blackbody model. (1) Observation; (2) neutral hydrogen column density from {\tt tbabs} model; (3) blackbody temperature; (4) blackbody radius; (5) inner disc temperature; (6) inner disc radius; (7) the unabsorbed total luminosity in the 0.3--10 keV band derived using {\tt cflux}; (8) ratio of the disc flux to the total flux; (9) $\chi^2$ statistics and degrees of freedom.}
 	\begin{tabular}{@{}ccccccccc@{}}
	\hline
	\hline
Obs & $N_{\rm H}$ & $\rm kT_{bb}$ &$\rm R_{bb}$ & $\rm kT_{in}$ &$\rm R_{in}$ & $\rm log L_{X}$ & $\rm Ratio$ & $\rm \chi^2/ d.o.f$ \\
& ($10^{22}~\rm cm^{-2}$) & (keV) & ($10^{3}$ km) & (keV) & ($10^{2}$ km) & ($\rm~erg~s^{-1}$) & & \\
\hline

$Swift$1 & $0.45^{+0.21}_{-0.21}$ & $0.11^{+0.02}_{-0.02}$ & $15.73^{+41.37}_{-11.72}$ & $0.58^{+0.26}_{-0.11}$ & $1.90^{+1.59}_{-1.11}$ & $39.54^{+0.67}_{-0.39}$ & $0.14$ & 45.9/62  \\
$Swift$2 & $0.53^{+0.20}_{-0.16}$ & $0.10^{+0.02}_{-0.02}$ & $30.93^{+54.75}_{-21.19}$ & $0.54^{+0.06}_{-0.06}$ & $2.83^{+1.27}_{-0.77}$ & $39.90^{+0.60}_{-0.42}$ & $0.10$ & 70.4/75  \\
$Swift$3 & $0.12^{+0.10}_{-0.12}$ & $0.19^{+0.09}_{-0.06}$ & $1.89^{+5.39}_{-1.14}$ & $0.75^{+0.21}_{-0.06}$ & $1.45^{+0.78}_{-0.83}$ & $39.19^{+0.24}_{-0.17}$ & $0.50$ & 62.4/70  \\

XMM1 & $0.20^{+0.01}_{-0.01}$ & $0.15^{+0.01}_{-0.01}$ & $3.93^{+0.49}_{-0.20}$ & $0.66^{+0.01}_{-0.01}$ & $1.27^{+0.08}_{-0.07}$ & $39.09^{+0.03}_{-0.02}$ & $0.29$ & 760.4/599  \\
XMM2 & $0.14^{+0.03}_{-0.01}$ & $0.18^{+0.02}_{-0.02}$ & $3.08^{+1.22}_{-0.75}$ & $0.71^{+0.03}_{-0.05}$ & $1.68^{+0.11}_{-0.12}$ & $39.32^{+0.10}_{-0.05}$ & $0.41$ & 401.2/368  \\
XMM3 & $0.22^{+0.06}_{-0.04}$ & $0.15^{+0.01}_{-0.01}$ & $6.53^{+3.36}_{-2.18}$ & $0.60^{+0.04}_{-0.04}$ & $2.71^{+0.52}_{-0.41}$ & $39.49^{+0.08}_{-0.08}$ & $0.35$ & 267.6/295  \\
XMM4 & $0.26^{+0.10}_{-0.08}$ & $0.14^{+0.02}_{-0.02}$ & $5.97^{+5.71}_{-2.80}$ & $0.59^{+0.07}_{-0.06}$ & $1.72^{+0.62}_{-0.45}$ & $39.22^{+0.28}_{-0.19}$ & $0.24$ & 203.6/169  \\
XMM5 & $0.19^{+0.10}_{-0.08}$ & $0.15^{+0.03}_{-0.02}$ & $4.89^{+4.76}_{-2.21}$ & $0.57^{+0.08}_{-0.06}$ & $2.22^{+0.89}_{-0.68}$ & $39.28^{+0.09}_{-0.17}$ & $0.31$ & 135.4/127  \\
XMM6 & $0.20^{+0.01}_{-0.01}$ & $0.16^{+0.01}_{-0.01}$ & $4.09^{+0.39}_{-0.36}$ & $0.63^{+0.01}_{-0.01}$ & $1.67^{+0.09}_{-0.08}$ & $39.21^{+0.01}_{-0.02}$ & $0.32$ & 868.3/687 \\

XMM6 (L) & $0.23^{+0.02}_{-0.02}$ & $0.15^{+0.01}_{-0.01}$ & $4.69^{+0.65}_{-0.61}$ & $0.63^{+0.02}_{-0.01}$ & $1.65^{+0.11}_{-0.11}$ & $39.21^{+0.02}_{-0.02}$ & $0.30$ & 594.7/476 \\
XMM6 (H) & $0.18^{+0.02}_{-0.01}$ & $0.17^{+0.01}_{-0.01}$ & $3.62^{+0.51}_{-0.42}$ & $0.64^{+0.02}_{-0.02}$ & $1.71^{+0.15}_{-0.13}$ & $39.22^{+0.02}_{-0.02}$ & $0.35$ & 528.7/462 \\

XMM2+N1 & $0.10^{+0.03}_{-0.03}$ & $0.21^{+0.02}_{-0.02}$ & $2.21^{+0.58}_{-0.39}$ & $0.83^{+0.06}_{-0.05}$ & $1.09^{+0.23}_{-0.19}$ & $39.26^{+0.05}_{-0.04}$ & $0.38$ &451.1/400 \\
XMM4+N2 & $0.26^{+0.10}_{-0.09}$ & $0.14^{+0.02}_{-0.02}$ & $5.86^{+5.56}_{-2.69}$ & $0.59^{+0.07}_{-0.06}$ & $1.72^{+0.61}_{-0.47}$ & $39.21^{+0.26}_{-0.20}$ & $0.25$ &216.1/177  \\
    
\hline
\end{tabular} 
\label{bbody_diskbb_params}
\end{table*}

For XMM1, XMM2, XMM4 and XMM6 observations, the doubly absorbed blackbody plus disc blackbody model provides a reduced $\chi{^2}$ of $\sim 1.1-1.3$. Thus, we added an optically thin thermal plasma model ({\tt APEC}) to the two-component model to check whether the additional model improves further the spectral fit. This model component improves the spectral fit in these observations compared to the blackbody plus disc blackbody model, and the $\Delta\chi{^2}$ is 7.7--87.1 with the loss of two additional d.o.f. The spectral parameters such as absorption, blackbody temperature and the inner disc temperature are similar to that obtained from the two-thermal component model, while the inferred plasma temperature is $\sim 0.2$ keV in these observations except XMM1 (see Table \ref{apec_edge_bbody_diskbb_params}). In XMM1 observation, the best-fit value of plasma temperature is $\sim 1.1$ keV. In addition, we replaced the {\tt APEC} with an absorption edge ({\tt edge} in {\sc xspec}), which improved the spectral fit significantly (see Table \ref{apec_edge_bbody_diskbb_params}) compared to the two-component model. The observed absorption edge is around 1 keV in these observations, and the absorption depth is in the range of 0.25--0.8.

\begin{table*}
\setlength{\tabcolsep}{10.0pt}
	\caption{Best-fit spectral parameters of additional spectral features fitted with optically-thin thermal plasma ({\tt apec}) or an absorption edge ({\tt edge}) along with blackbody plus disc blackbody model. (1) Observation; (2) plasma temperature; (3) {\tt apec} normalization; (4) difference in $\rm \chi^2$ when the {\tt apec} model added to doubly absorbed blackbody plus disc blackbody model; (5) $F$-test significance; (6) threshold energy; (7) absorption depth at the threshold; (8) difference in $\rm \chi^2$ when the {\tt edge} model added to doubly absorbed blackbody plus disc blackbody model; (9) $F$-test significance.}
 	\begin{tabular}{@{}ccccccccc@{}}
	\hline
	\hline
Obs & $\rm kT_{apec}$ &$\rm N_{apec}$ & $\Delta\chi{^2}_{apec}$ & Significance & $E$ & $\rm \tau_{max}$ & $\Delta\chi{^2}_{edge}$ & Significance\\
 & (keV) & ($10^{-4}$) & & & (keV) & & & \\
\hline

XMM1 & $1.08^{+0.11}_{-0.04}$ & $0.47^{+0.13}_{-0.04}$ & 87.1 & $>99.9\%$ & $1.13^{+0.01}_{-0.01}$ & $0.42^{+0.03}_{-0.06}$ & 104.0 & $>99.9\%$ \\
XMM2 & $0.23^{+0.05}_{-0.04}$ & $3.37^{+3.28}_{-1.75}$ & 12.1 & $>99\%$ & $1.12^{+0.10}_{-0.11}$ & $0.25^{+0.16}_{-0.17}$ & 5.6 & $>92\%$ \\
XMM3 & $0.17^{+0.17}_{-0.10}$ & $5.52^{+17.0}_{-5.52}$ & 2.7 & $>77\%$ & $1.34^{+0.13}_{-0.10}$ & $0.63^{+2.48}_{-0.32}$ & 3.5 & $>85\%$ \\
XMM4 & $0.16^{+0.05}_{-0.03}$ & $16.6^{+101.0}_{-13.1}$ & 7.7 & $>60\%$ & $1.12^{+0.03}_{-0.02}$ & $0.81^{+0.41}_{-0.25}$ & 16.7 & $>99.9\%$ \\
XMM5 & $0.04^{+0.21}_{-0.04}$ & $<0.01$ & 0.7 & $>27\%$ & $0.70^{+0.04}_{-0.03}$ & $0.44^{+0.34}_{-0.24}$ & 4.5 & $>87\%$ \\
XMM6 & $0.17^{+0.01}_{-0.01}$ & $3.96^{+1.43}_{-1.30}$ & 60.0 & $>99.9\%$ & $1.20^{+0.03}_{-0.02}$ & $0.38^{+0.07}_{-0.07}$ & 81.7 & $>99.9\%$ \\

\hline
\end{tabular} 
\label{apec_edge_bbody_diskbb_params}
\end{table*}

\begin{figure*}

 \includegraphics[width=8.2cm,angle=0]{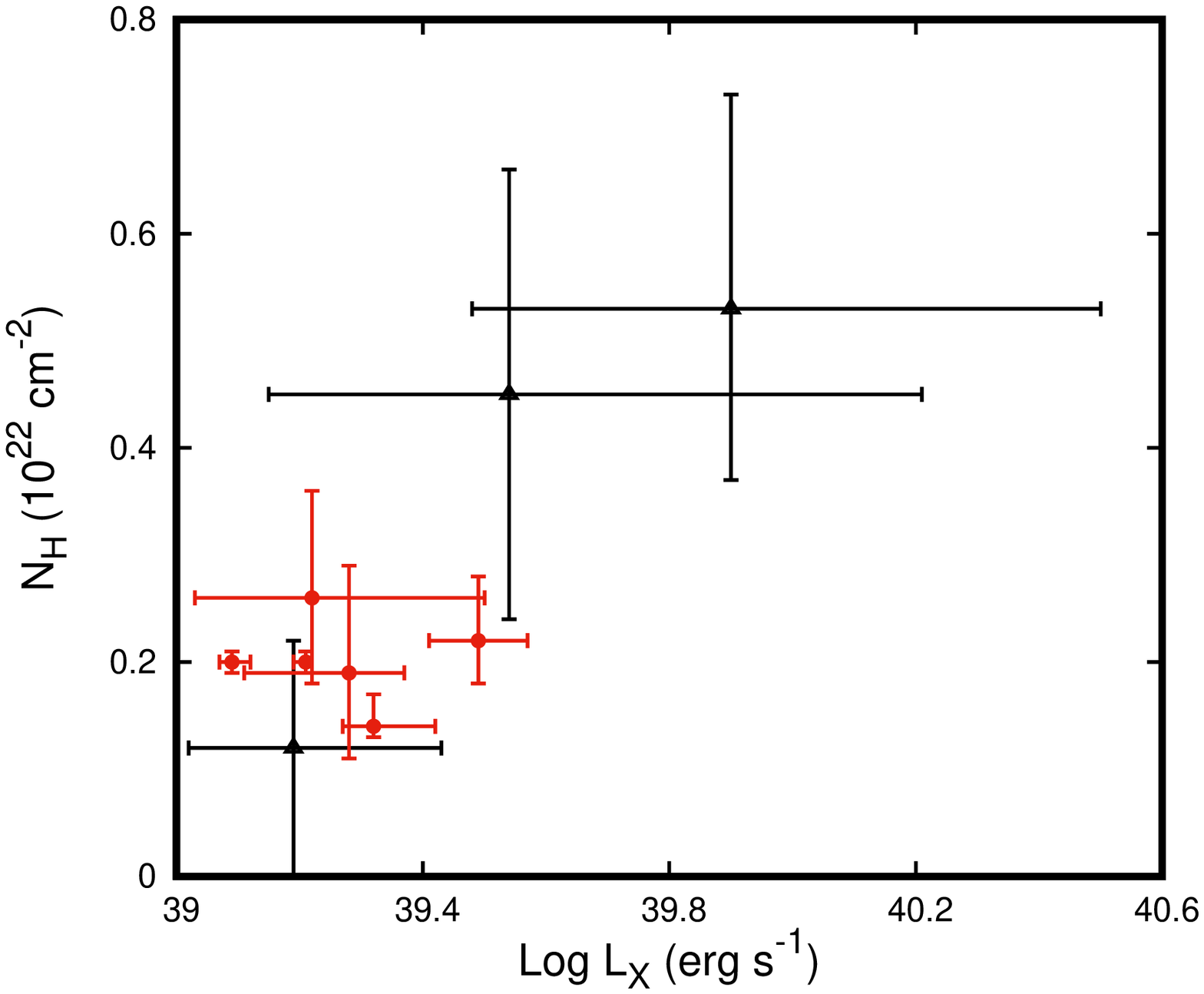}
 \includegraphics[width=8.2cm,angle=0]{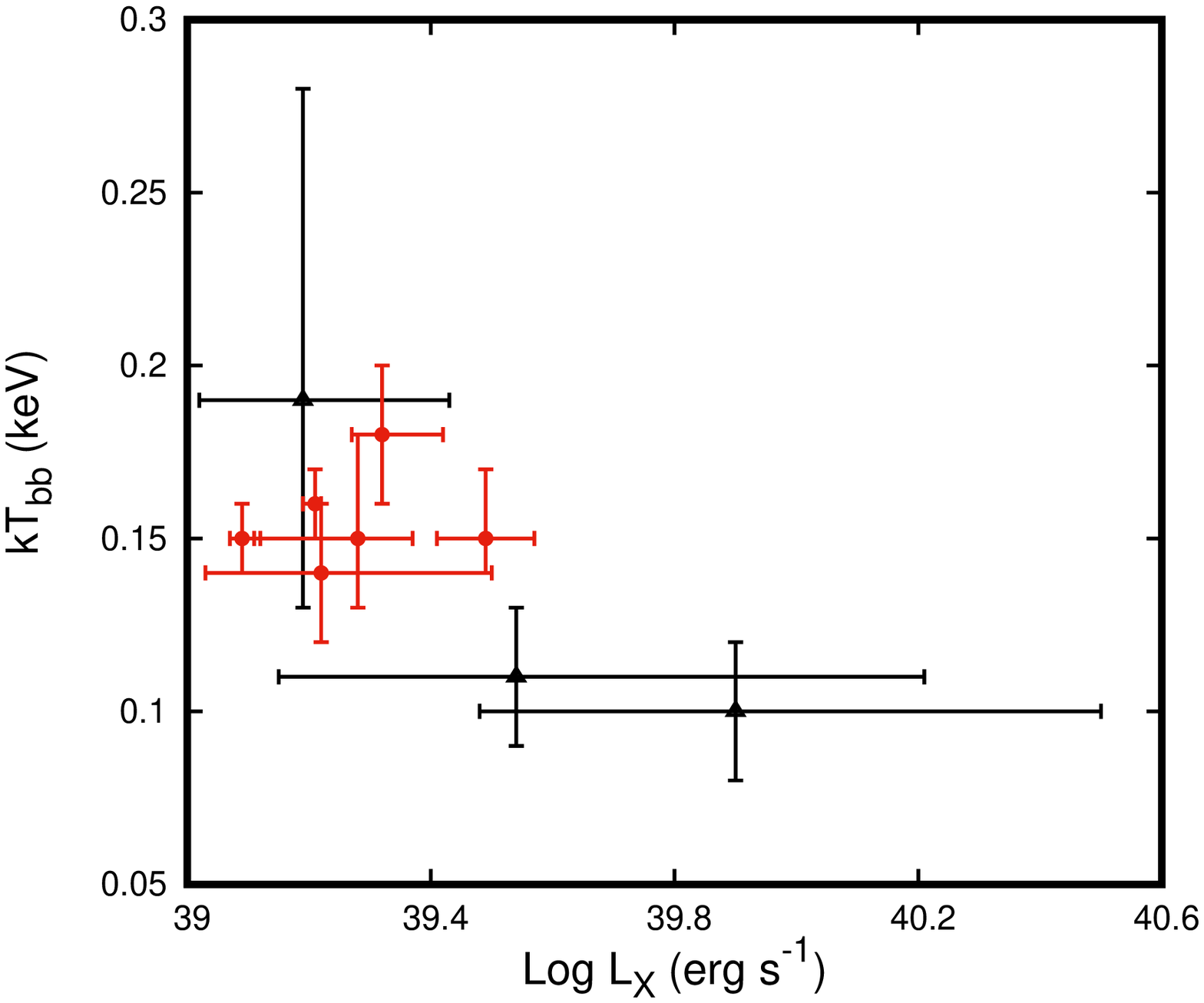}

 \includegraphics[width=8.2cm,angle=0]{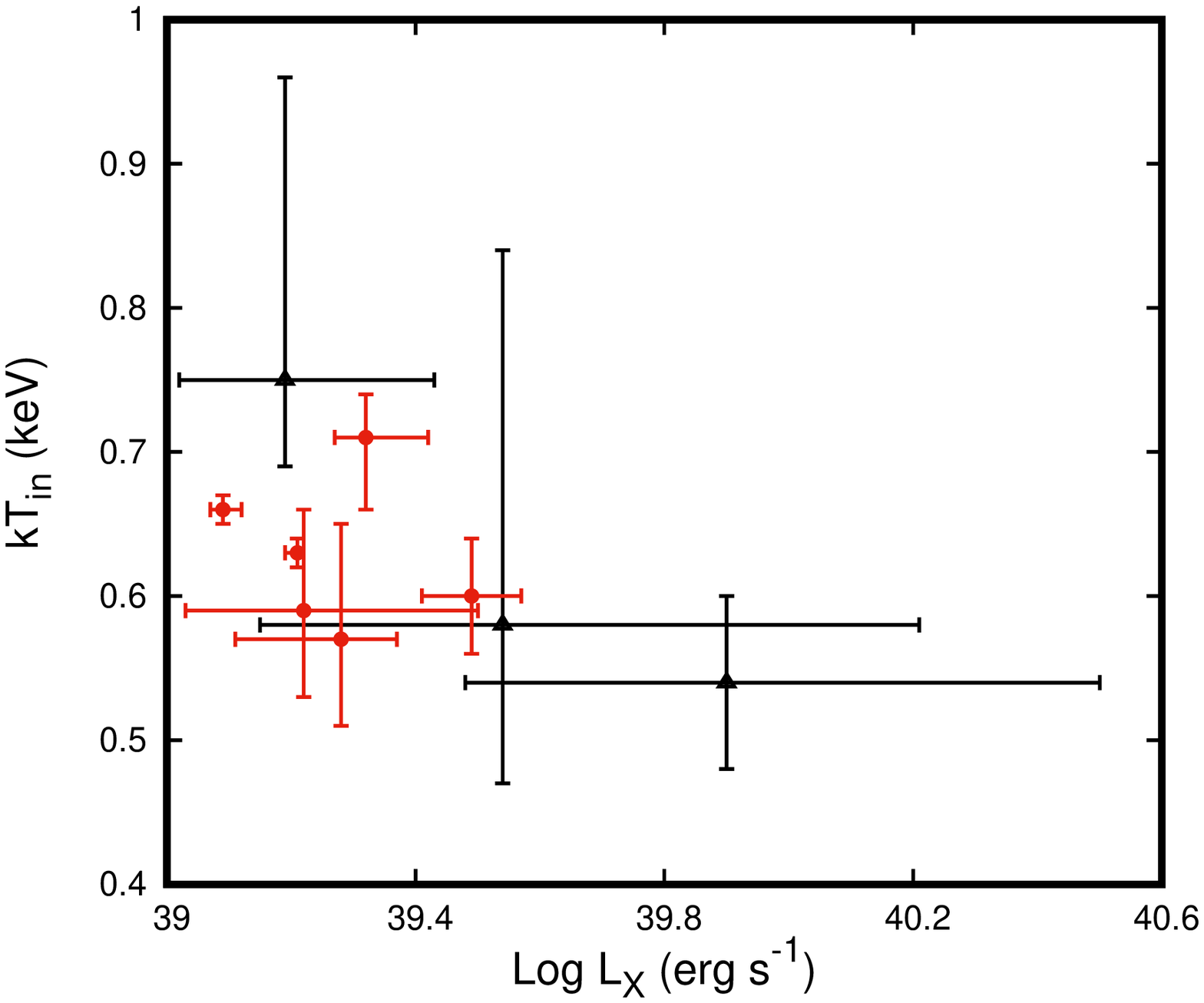}
 \includegraphics[width=8.2cm,angle=0]{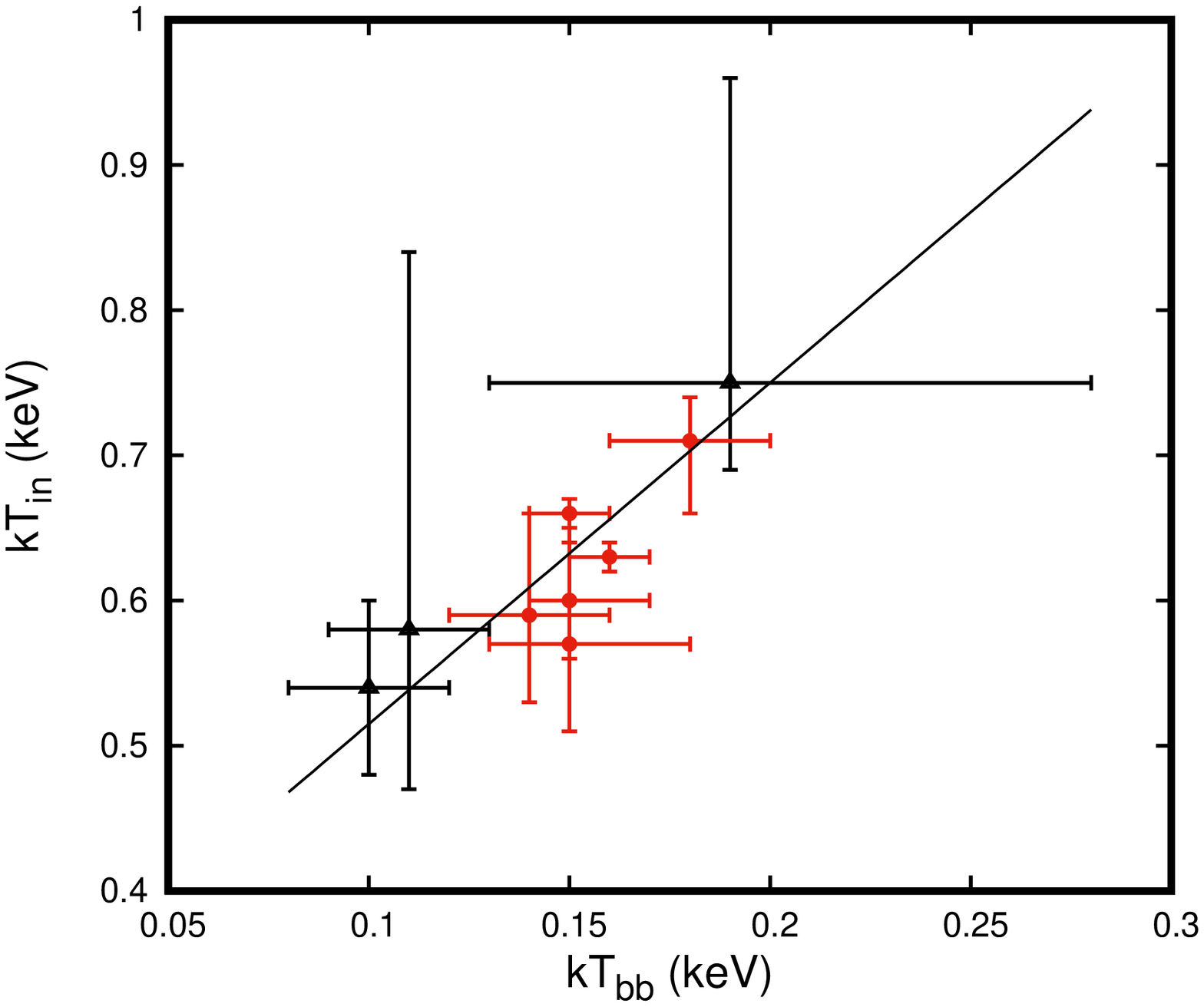}

 \includegraphics[width=8.2cm,angle=0]{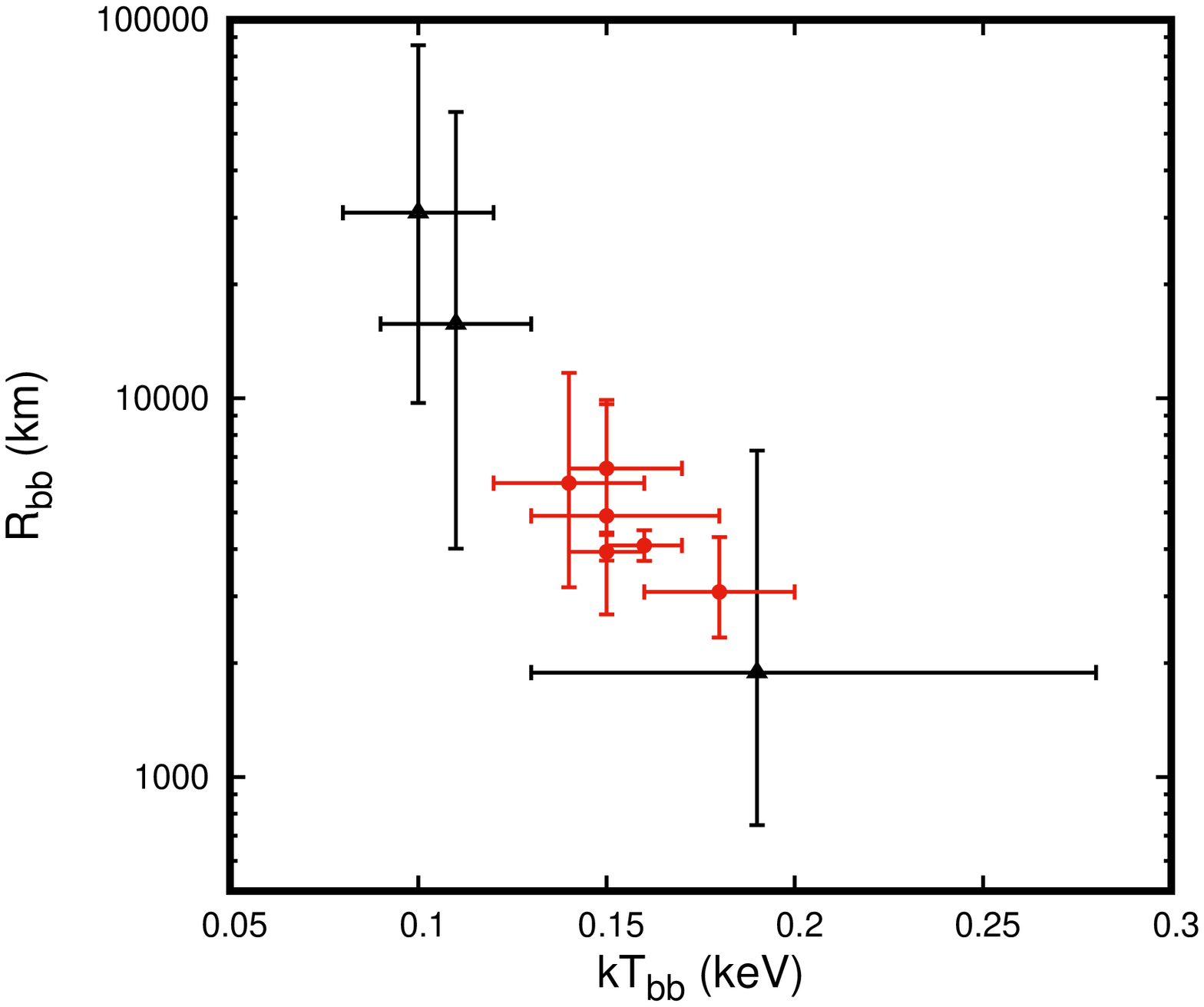}
 \includegraphics[width=8.2cm,angle=0]{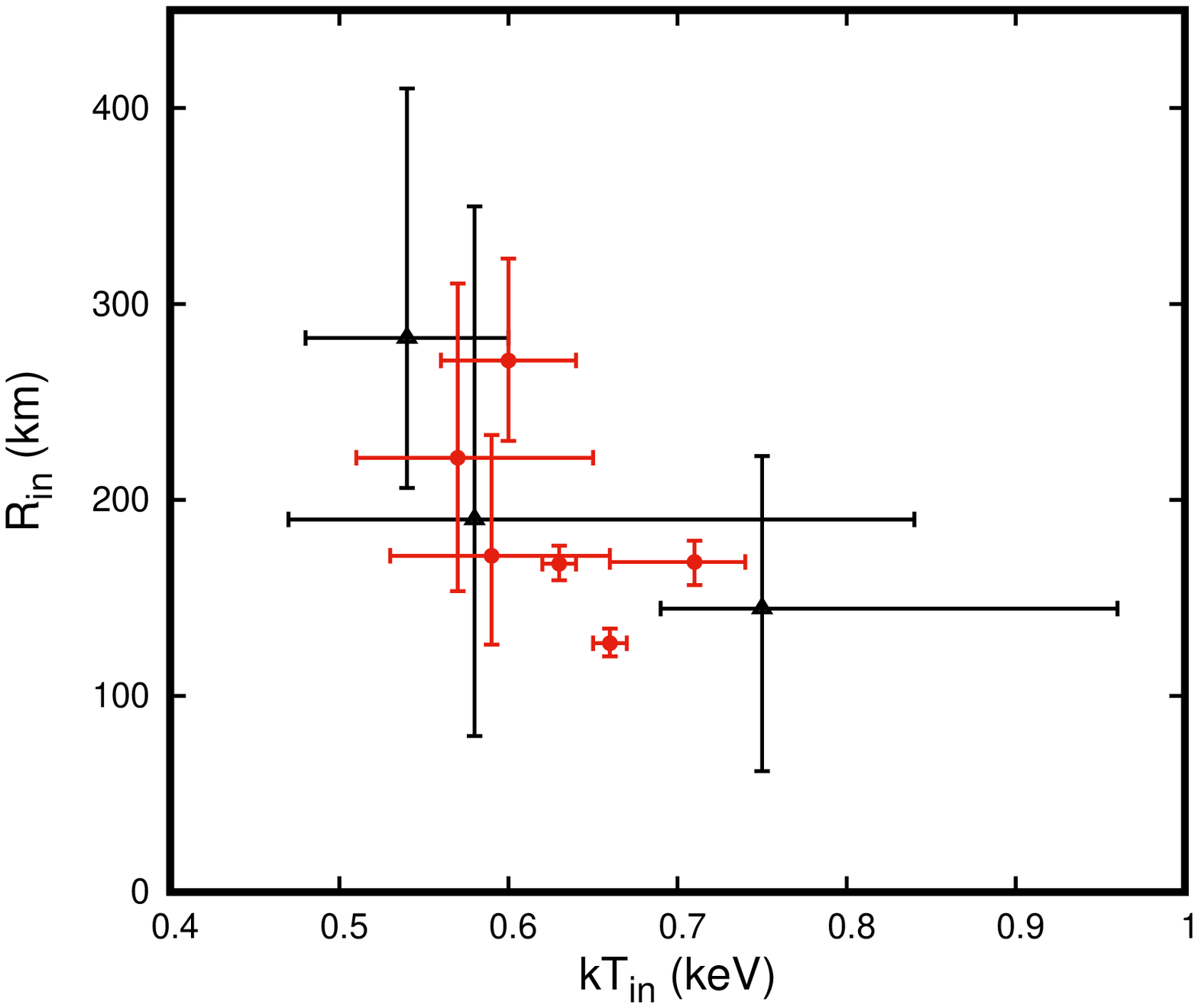}

\caption{Variations of best-fit spectral parameters obtained from the doubly absorbed blackbody and disc blackbody model. The black triangles and red circles represent the {\it Swift} and {\it XMM-Newton} data, respectively. The solid line in the $\rm kT_{bb}$ vs $\rm kT_{in}$ plot represents the correlation between the parameters derived from the {\sc linmix} method.}
\label{spec_params}
\end{figure*}

In {\it XMM-Newton} observations, the source exhibits variation in the HR as the intensity increases, while we do not observe a clear variation in the {\it Swift} observations. Thus, we further investigate the source variability using spectral parameters as a function of the unabsorbed luminosity derived from the two-thermal component model. The spectral parameters variations are shown in Figure \ref{spec_params}. From these plots, it is clear that the blackbody temperature and inner disc temperature show a decreasing trend with the luminosity, while the absorption column density increases as the luminosity increases. However, the trend is mainly driven by the {\it Swift} data which have a large uncertainty on all parameters, including the luminosity. If we do not consider the {\it Swift} data, we observe a marginal change in spectral parameters and luminosity (by a factor of $\sim 2.6$). The disc temperature varies with the blackbody temperature and follows a linear correlation, $\rm kT_{disc} = (1.71 \pm 0.52) kT_{bb} + (0.37 \pm 0.08)$. We used the non-linear least-squares (NLLS) Marquardt-Levenberg algorithm to derive the correlation. In addition, we used a linear regression based on the Bayesian approach called {\sc linmix}\footnote{\url{https://linmix.readthedocs.io/en/latest/}} \citep{Kel07}, which considers the measurement errors on both parameters. In this method, the temperatures follow a relation, $\rm kT_{disc} = (2.35 \pm 6.12) kT_{bb} + (0.28 \pm 0.92)$ (see the middle right panel of Figure \ref{spec_params}). For this correlation, we calculated the Spearman rank correlation coefficient ($r_{s}$) and the probability ($p$-value) for the null hypothesis, which are 0.87 and $2.63 \times 10^{-3}$, respectively. We also calculated the blackbody radius and inner disc radius from the normalization of the respective models. We observed a correlation between the blackbody temperature and radius, which follow $R_{bb} \propto T_{bb}^{-4.5}$ relation ($r_{s} = -0.92$, $p$-value = $5.34 \times 10^{-4}$), but if we consider {\it XMM-Newton} data alone, the relation becomes $R_{bb} \propto T_{bb}^{-2.6}$ ($r_{s} = -0.70$, $p$-value = 0.12). Thus, it is clear that the source showed a negative correlation between the blackbody temperature and radius. 

\begin{figure*}
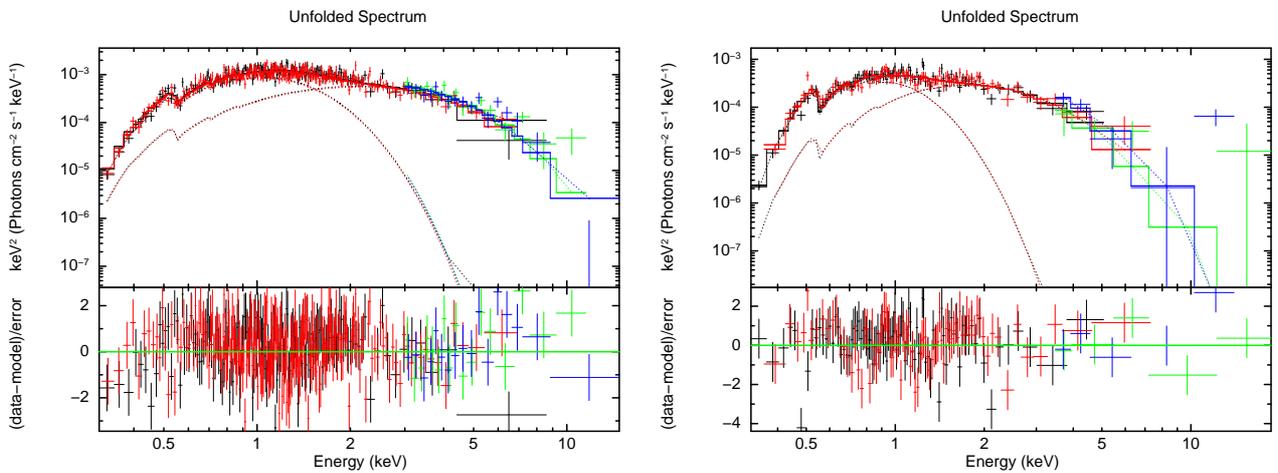


 \includegraphics[width=6.2cm,angle=-90]{xmm2_nustar1_2tbabs_bbody_diskbb.ps}
 \includegraphics[width=6.2cm,angle=-90]{xmm4_nustar2_2tbabs_bbody_diskbb.ps}

\caption{The 0.3--20 keV broadband X-ray spectrum of NGC 55 ULX1. The spectra from the simultaneous {\it XMM-Newton} (XMM2) \& {\it NuSTAR} (N1) (left panel) and XMM4 \& N2 (right panel) observations fitted with doubly absorbed blackbody and disc blackbody components. The black, red, green and blue data points represent {\it XMM-Newton} PN, MOS, {\it NuSTAR} FPMA and FPMB, respectively.}

\label{xmm_nustar}
\end{figure*}

We also investigated the broadband spectral properties of NGC 55 ULX1 using the simultaneous {\it XMM-Newton} and {\it NuSTAR} observations conducted in 2019 and 2020. For the simultaneous observations XMM2 \& N1 and XMM4 \& N2, we performed the broadband X-ray spectral modelling using EPIC PN, MOS, {\it NuSTAR} FPMA and FPMB data in the 0.3--20 keV energy band. Unlike other ULXs \citep{Wal14, Muk15, Ran15}, NGC 55 ULX1 has no or little emission above the 10 keV energy. Thus, we used the two-thermal component model for the broadband spectrum, and the model provided an acceptable fit (see Figure \ref{xmm_nustar}). The best-fit spectral parameters are marginally varied between the two epochs but consistent with uncertainties, and parameters are given in Table \ref{bbody_diskbb_params}. In addition, we applied the accreting magnetic NS continuum model consists of a power-law with a high-energy exponential cut-off plus a blackbody ({\tt bbody + highecut $\times$ PL} in {\sc xspec}) to describe the simultaneous broadband {\it XMM-Newton} and {\it NuSTAR} data. This model is widely used to describe the X-ray spectrum of Galactic X-ray pulsars and later applied to ULX spectra to search for ULX pulsars \citep[][and references therein]{Pin17a}. The model provided a satisfactory fit in both the epochs and yielded parameters given in Table \ref{pulsator_model}. The spectral parameters, in particular $E_{c}$ and $E_{f}$, are different from that of pulsating ULXs and candidate ULX pulsars \citep{Pin17a, Jit20}. We also derived the softness and hardness parameters using the flux derived from the ``pulsator-like'' model. The softness is defined as the ratio of flux in the 2--4 and 4--6 keV bands, while the hardness is defined as the ratio of flux in the 6--30 and 4--6 keV bands. The derived softness and hardness parameters are in the range of 5.1--9.8 and 0.1--0.3, respectively.  

\begin{table*}
\setlength{\tabcolsep}{2.0pt}
	\caption{The X-ray spectral parameters for NGC~55~ULX1 from the ``pulsator-like'' model. (1) Observation; (2) neutral hydrogen column density from {\tt tbabs} model; (3) blackbody temperature; (4) blackbody  normalization; (5-6) cut-off energy and e-folding energy; (7) PL index; (8) PL normalization; (9) softness; (10) hardness; (11) the unabsorbed total luminosity in the 0.3--20 keV band derived using {\tt cflux}; (12) $\chi^2$ statistics and degrees of freedom.}
 	\begin{tabular}{@{}cccccccccccc@{}}
	\hline
	\hline
Obs & $N_{\rm H}$ & $\rm kT_{bb}$ &$\rm N_{bbody}$ & $\rm E_{c}$ & $\rm E_{f}$ & $\Gamma$ & $\rm N_{PL}$ & $\rm Softness$ & $\rm Hardness$ & $\rm log L_{X}$ & $\rm \chi^2/ d.o.f$ \\
 & ($10^{22}~\rm cm^{-2}$) & (keV) & $(10^{-4})$ & (keV) & (keV) & & $(10^{-3})$ & & & ($\rm~erg~s^{-1}$) & \\

\hline

XMM2+N1 & $0.17^{+0.10}_{-0.06}$ & $0.18^{+0.02}_{-0.03}$ & $0.52^{+0.29}_{-0.10}$ & $1.75^{+0.14}_{-0.21}$ & $1.53^{+0.67}_{-0.37}$ & $1.55^{+0.76}_{-0.83}$ & $0.69^{+0.60}_{-0.34}$ & $5.13 \pm 0.67$ & $0.29 \pm 0.09$ & $39.40^{+0.19}_{-0.11}$ & $428.9/398$ \\
XMM4+N2 & $0.24^{+0.11}_{-0.09}$ & $0.15^{+0.03}_{-0.03}$ & $0.44^{+0.55}_{-0.18}$ & $1.67^{+0.16}_{-0.15}$ & $0.78^{+0.62}_{-0.31}$ & $0.49^{+1.58}_{-2.32}$ & $0.17^{+0.27}_{-0.13}$ & $9.79 \pm 4.30$ & $0.10 \pm 0.14$ & $39.14^{+0.31}_{-0.18}$ & $206.2/175$ \\

\hline
\end{tabular} 
\label{pulsator_model}
\end{table*}

\section{Discussion}
\label{sec:discu}

In this work, we have studied the spectral and temporal properties of NGC 55 ULX1 using {\it Swift}, {\it XMM-Newton} and {\it NuSTAR} observations conducted during 2013-2021. The source exhibited variability during these observations, where the flux varied by a factor of $\sim$ 5--6. We used the two-thermal component ({\tt bbody} and {\tt diskbb}) model to describe the observed spectra, and spectral parameters such as absorption column density, blackbody and inner disc temperatures showed a marginal trend with X-ray luminosity as the source evolved. In particular, we observed a strong correlation between the blackbody and inner disc temperatures. The correlation is consistent with previous studies using the {\it Swift} and {\it Chandra} observations \citep{Pin15}, which suggests that both components are co-evolving with each other. In addition, the characteristic radius derived from the blackbody component is anti-correlated with the blackbody temperature as seen in ULSs \citep{Urq16, Sor16}. 

NGC 55 ULX1 identified in the SUL state in most of the previous works \citep{Gla09, Sut13} and is a good candidate for a transitional object between classical ULXs and ULSs \citep{Pin17}. Modelling the NGC 55 ULX1 spectra with MCD plus power law model suggests that the source spectrum is very soft ($\Gamma > 3$) similar to ULSs, but a bright hard tail above 1 keV can be seen in these spectra. If we consider the supersoft source criteria ($(M-S)/T \lesssim -0.8$, where M, S and T are the count rates in the 1.1--2.5, 0.3--1.1 keV and 0.3--7 keV energy bands) used in \citet{Urq16}, NGC 55 ULX1 does not belong to the ULS category ($(M-S)/T$ is in the range of $\sim -0.30$ -- $-0.15$). In addition, the hard component in the blackbody plus disc blackbody model contributing more than 20 per cent in the total spectrum of NGC 55 ULX1 in the {\it XMM-Newton} observations, however for classical supersoft sources, the fraction of the energy carried by the photons above 1.5 keV is slightly above 10\%. Thus, our analysis confirmed that the source stayed in the SUL state in these observations and did not make a transition to the ULS state, even though the flux varied by a factor of $\sim 6$. 

Although the source is in the SUL state, the soft component properties are similar to that of ULSs. In ULSs, multiple origins have been proposed for the soft emission, which includes disc dominated intermediate-mass black hole, quasi-steady nuclear burning on the surface of white dwarfs in close binary systems and outflows driven by super-Eddington accreting stellar-mass black holes or neutron stars \citep{Urq16, Sor16}. The characteristic radius and temperature inferred from the modelling of NGC 55 ULX1 spectra are inconsistent with the first two scenarios. In addition, we do not find an L $\propto T^{4}$ relation for NGC 55 ULX1, which rules out the emission from a cool disc around a more massive black hole. Given the negative R--T correlation, a promising explanation is the outflows due to supercritical accretion. The absorption/emission lines \citep{Sut15, Pin16, Pin17, Kos21} and soft X-ray spectral residuals \citep{Mid14, Mid15, Urq16, Sor16, Fen16} detected in the ULXs spectra are considered as the signatures of outflow. In the high-quality {\it XMM-Newton} spectra of NGC 55 ULX1, we observed the absorption edge near 1 keV, which may be due to the L-edge absorption by highly ionized iron in the clumpy winds. In addition to the dominant soft component, we detected soft residuals (particularly in the 0.7--2 keV band) in some {\it XMM-Newton} observations. This feature can be modelled with an optically thin thermal plasma model and the addition of this component showed significant improvement (at a significance level $>99\%$) in the fit in some observations compared to the two-thermal component model. However, the obtained plasma temperature for NGC 55 ULX1 is $\sim 0.2$ keV, which is smaller than that of other soft ULXs \citep{Urq16, Sor16, Fen16}. This feature is due to either the broadened, blue-shifted absorption by a partially ionized, optically thin outflow or the emission by collisionally ionized gas \citep{Mid14, Mid15}. Thus, the observed spectral features strongly support the supercritical radiatively driven outflow scenario for NGC 55 ULX1. 

The hard emission component is detected in all the spectra of NGC 55 ULX1, and we modelled this component using the disc blackbody model. \citet{Fen16} interpreted the hard component as the leaked emission from the inner disc via the central funnel or advected through the wind. This argument may be true for NGC 55 ULX1 and is consistent with the co-evolution of soft and hard components, which is inferred from the positive correlation between blackbody and inner disc temperatures. In ULSs, the appearance of the hard component is related to the blackbody radius and the blackbody temperature. When the blackbody radius is smallest, say $r_{bb} \lesssim$ 20000 km, and the blackbody temperature is highest, say $T_{bb} \gtrsim$ 0.1 keV, the photosphere reduces its size, and a more direct view of the inner region is possible \citep{Urq16}. Although NGC 55 ULX1 in the SUL state, the characteristic radius and temperature derived from the blackbody component is generally consistent with the criteria listed above. It suggests that the hard component originates from the inner regions of accretion flow.

Using the simultaneous {\it XMM-Newton} and {\it NuSTAR} observations, we explored the broadband properties of NGC 55 ULX1. The hard tail above 10 keV energy has been detected in several ULXs \citep{Wal14, Muk15, Ran15}, while we did not detect this hard tail in NGC 55 ULX1. We investigated the temporal properties of the source using {\it NuSTAR} and {\it XMM-Newton} observations but failed to identify the pulsation from the source. Alternatively, we used the broadband spectral modelling with the ``pulsator-like'' model to explore the pulsating nature of the source. The spectral parameters inferred from the accreting magnetic NS continuum model are not similar to that of pulsating ULXs, and the X-ray colours derived from the ``pulsator-like'' model are extremely soft, which disfavour NS compact object for NGC 55 ULX1. Furthermore, the characteristic radius derived from the blackbody component of NGC 55 ULX1 varies between the observations, which does not favour an NS as the origin of the thermal emission. 

The X-ray flux dipping episodes have been observed in many ULXs, for example NGC 55 ULX1 \citep{Sto04, Pin15}, NGC 628  \citep{Liu05}, M94 \citep{Lin13}, NGC 5408 \citep{Pas13, Gri13}, NGC 247 \citep{Fen16, Ai21, Als21, Pin21}, M 51 ULX--7 \citep{Vas21, Hu21}, M51 and M81 \citep{Urq16}. Significant X-ray dipping behaviour has been reported in NGC 55 ULX1 using the {\it XMM-Newton} observations conducted in 2001 \citep{Sto04}. However, marginal hints of a limited number of dips detected in the later {\it Swift} and {\it Chandra} observations \citep{Pin15}. We also detected marginal evidence of dipping events in the  {\it Swift} and {\it XMM-Newton} observations considered in this work, and the flux dropped by $\sim 5-73\%$ from the average value in the 0.3--10 keV band. However, if we consider the hard band, the depth of dips increases to somewhat higher values, which is similar to the behaviour observed in the {\it XMM-Newton} observations conducted in 2001 \citep{Sto04}. Although the depth of the dips is not strong as seen in 2001 {\it XMM-Newton} observations, we observed the probable dips in different intensity levels, which suggests that the dips are not associated with any particular flux levels. Multiple origins have been proposed for dips in ULXs, which include occultation of obscuring material, propeller scenario and outflows driven by super-Eddington accretion \citep{Sto04, Fen16, Ai21, Als21}. However, its origin is still debated. In NGC 55 ULX1, the observed dips may be due to changes in the photosphere of the outflow, which results from the local change in accretion rate or the thermal instabilities in the wind or disc. The wind origin of the dips is further supported by the properties (temperature and emitting size) of the blackbody component.

In summary, we studied the spectral and temporal properties of NGC 55 ULX1 using {\it Swift}, {\it XMM-Newton} and {\it NuSTAR} observations conducted during 2013--2021. The source was identified to be in the $soft~ultraluminous$ state of ULXs in these observations. We observed a positive correlation between the blackbody and inner disc temperatures, while the characteristic radius and blackbody temperature are negatively correlated when X-ray spectra are fitted with the two-thermal component model. We observed marginal hints of dip episodes in these observations, however, they are not very strong as seen in the previous {\it XMM-Newton} observation. A supercritical radiatively driven outflow scenario can explain the observed properties of the source. The future observations with {\it XMM-Newton} and {\it NuSTAR} in different spectral states can shed further light on the nature of the source.

\section*{Acknowledgements}
The author thanks the anonymous referee for the constructive comments and suggestions that improved this manuscript. The author thanks Ranjeev Misra, C. D. Ravikumar and Zahir Shah for the useful discussion and the timely help. This research has made use of {\it Swift}, {\it XMM-Newton} and {\it NuSTAR} data provided by the High Energy Astrophysics Science Archive Research Center (HEASARC), which is a service of the Astrophysics Science Division at NASA/GSFC. This research has made use of the XRT Data Analysis Software (XRTDAS) developed under the responsibility of the ASI Science Data Center (ASDC), Italy and NuSTAR Data Analysis Software (NuSTARDAS) jointly developed by the ASI Science Data Center (ASDC, Italy) and the California Institute of Technology (Caltech, USA).\\


\section*{Data Availability}

The data used in this article are available in the HEASARC database ({\url{https://heasarc.gsfc.nasa.gov}). 








\bsp	
\label{lastpage}
\end{document}